\documentclass{amsart}
\usepackage{amsmath}
\usepackage{amsthm}
\usepackage{amsfonts}
\usepackage{amssymb}
\usepackage{amsbsy}
\usepackage[dvips]{graphicx}
\usepackage{youngtab}
\usepackage{amsaddr}

\newtheorem{theorem}{Theorem}
\newtheorem{proposition}[theorem]{Proposition}

\theoremstyle{definition}
\newtheorem{problem}{Problem}

\newtheorem{remark}{Remark}

\newcommand{\beq}{\begin{equation}}
\newcommand{\eeq}{\end{equation}}

\newcommand{\hsps}{\hspace{0.5cm}}

\newcommand{\Real}{\mathop{\mathrm{Re}}}
\newcommand{\Imag}{\mathop{\mathrm{Im}}}

\newcommand{\R}{\mathbb{R}}
\newcommand{\C}{\mathbb{C}}
\newcommand{\HH}{\mathbb{H}}
\newcommand{\SSS}{\mathbb{S}}

\newcommand{\fsu}{\mathfrak{su}}
\newcommand{\fso}{\mathfrak{so}}
\newcommand{\fsp}{\mathfrak{sp}}
\newcommand{\fsl}{\mathfrak{sl}}

\newcommand{\fu}{\mathfrak{u}}
\newcommand{\fg}{\mathfrak{g}}

\newcommand{\SU}{\mathop{\mathrm{SU}}}
\newcommand{\SO}{\mathop{\mathrm{SO}}}
\newcommand{\OO}{\mathop{\mathrm{O\,\!}}}
\newcommand{\rU}{\mathop{\mathrm{U\,\!}}}
\newcommand{\Sp}{\mathop{\mathrm{Sp}}}
\newcommand{\SL}{\mathop{\mathrm{SL}}}
\newcommand{\GL}{\mathop{\mathrm{GL}}}
\newcommand{\SSSS}{\mathop{\mathrm{S}}}
\newcommand{\GG}{\mathop{\mathrm{G}}}

\newcommand{\cZ}{\mathcal{Z}}
\newcommand{\cU}{\mathcal{U}}
\newcommand{\cH}{\mathcal{H}}

\newcommand{\bxi}{\boldsymbol{\xi}}
\newcommand{\bXi}{\boldsymbol{\Xi}}
\newcommand{\bK}{\boldsymbol{K}}

\newcommand{\bbeta}{\boldsymbol{\eta}}
\newcommand{\bPi}{\boldsymbol{\Pi}}
\newcommand{\bGamma}{\boldsymbol{\Gamma}}

\newcommand{\bR}{\mathbf{R}}
\newcommand{\bZ}{\mathbf{Z}}
\newcommand{\bA}{\mathbf{A}}
\newcommand{\bB}{\mathbf{B}}
\newcommand{\bL}{\mathbf{L}}

\newcommand{\br}{\mathbf{r}}

\newcommand{\bp}{\mathbf{p}}
\newcommand{\bq}{\mathbf{q}}

\newcommand{\rmd}{\mathrm {d}}
\newcommand{\rmi}{\mathrm {i}}
\newcommand{\rmj}{\mathrm {j}}
\newcommand{\Tr}{\mathrm {Tr}}

\begin{document}

\title[Transition from resonances to surface waves]
{Transition from resonances to surface waves in $\boldsymbol{\pi^+}$--p elastic scattering}

\author[E. De Micheli]{Enrico De Micheli}
\address{\sl IBF -- Consiglio Nazionale delle Ricerche \\ Via De Marini, 6 - 16149 Genova, Italy}
\email{enrico.demicheli@cnr.it}

\author[G. A. Viano]{Giovanni Alberto Viano}
\address{\sl Dipartimento di Fisica -- Universit\`a di Genova,\\
Istituto Nazionale di Fisica Nucleare -- Sezione di Genova, \\
Via Dodecaneso, 33 - 16146 Genova, Italy}
\email{viano@ge.infn.it}

\begin{abstract}
In this article we study resonances and surface waves in $\pi^+$--p scattering.
We focus on the sequence whose spin--parity values are given by
$J^p=\frac{3}{2}^+,\frac{7}{2}^+,\frac{11}{2}^+,\frac{15}{2}^+,\frac{19}{2}^+$.
A widely--held belief takes for granted that this sequence
can be connected by a moving pole in the complex angular momentum (CAM) plane,
which gives rise to a linear trajectory of the form $J=\alpha_0+\alpha' m^2$, $\alpha'\sim 1/(\mathrm{GeV})^2$,
which is the standard expression of the Regge pole trajectory. But
the phenomenology shows that only the first few resonances
lie on a trajectory of this type. For
higher $J^p$ this rule is violated and is substituted by
the relation $J\sim kR$, where $k$ is the pion--nucleon c.m.s. momentum,
and $R\sim 1$ fm. In this article we prove:
(a) Starting from a non--relativistic model of the proton, regarded as composed by
three quarks confined by harmonic potentials, we prove that the first three members
of this $\pi^+$--p resonance sequence can be associated with a vibrational spectrum
of the proton generated by an algebra $\fsp(3,\R)$. Accordingly, these first three
members of the sequence can be described by Regge poles and lie on
a standard linear trajectory.
(b) At higher energies the amplitudes are dominated by diffractive scattering,
and the creeping waves play a dominant role. They can be described
by a second class of poles, which can be called Sommerfeld's poles,
and lie on a line nearly parallel to the imaginary axis of the CAM--plane.
(c) The Sommerfeld pole which is closest to the real axis of the CAM--plane is
dominant at large angles, and describes in a proper way the backward diffractive
peak in both the following cases: at fixed $k$, as a function of the scattering
angle, and at fixed scattering angle $\theta=\pi$, as a function of $k$.
(d) The evolution of this pole, as a function of $k$, is given
in first approximation by $J\simeq kR$.
\end{abstract}

\maketitle

\newpage

\section{Introduction}
\label{se:introduction} At the end of their review paper on baryon
spectroscopy, Hey and Kelly write \cite{Hey}: ``\emph{The ideas of spin
$\frac{1}{2}$ quarks and a hidden color degree of freedom must
surely rate as the most significant achievements of baryon
spectroscopy. Another piece of current dogma, taking much support
from the baryon spectrum, is the widely--held belief in linear
Regge trajectories as a function of mass--squared}''. But, after a
more careful phenomenological analysis, based, in particular, on
the work of Hendry \cite{Hendry1}, they conclude: ``\emph{The conventional
picture of linear Regge trajectories with universal slope is not
well--established.}'' Regarding the $\pi^+$--p collision, Hendry
writes \cite{Hendry2}: ``\emph{The first few resonances are consistent with a
straight--line trajectory; however, as we increase the spins, the
resonances appear to deviate from this trajectory.}''

In the conventional Regge--type phenomenology, the relationship
between the total spin $J$ and the squared--mass is given (with
standard notation) by: $J=\alpha_0+\alpha' m^2$, the slope
of the trajectory being $\alpha'\sim 1/(\mathrm{GeV})^2$.

Furthermore, it is often found in the literature the expression ``rotational excitations''
in view of the fact that states of the hadronic spectrum lying on the
same trajectory possess the property $\Delta J = 2$ \cite[pag. 4]{Barger}\cite{Dalitz}.
In this connection several authors, notably
Dothan, Gell--Mann, Ne'eman \cite{Dothan}, \v{S}ija\v{c}ki \cite{Sijacki},
and Ne'eman \cite{Neeman}, call the attention to the non--compact algebra
of the $\SL(3,\R)$ group which, on the other hand, plays a relevant
role in describing the nuclear rotational motion \cite{Sijacki}.
Therefore, we can say that the scenario appears to be far from being
neat and clear, nor it is changed significantly in more recent time.

In this article, instead of studying hadronic sequences in their
wide generality, we focus only on the sequence of resonances
obtained in the $\pi^+$--p elastic scattering, whose $J^p$ values
are given by:
$J^p=\frac{3}{2}^+,\frac{7}{2}^+,\frac{11}{2}^+,(\frac{15}{2}^+,\frac{19}{2}^+)$.
This is one of the most widely explored sequences in particle
physics, and is generated by the interaction of a $\pi^+$ meson with a proton:
if the total angular momentum of the proton (composed of three quarks) is $L=0$, then
adding the angular momentum of the pion, which is $1\hbar$ ($\hbar=1$),
and the proton spin, we have $J^P=\frac{3}{2}^+$.
If we consider also the isospin of the $\pi^+$--p system,
we have the famous $\Delta(\frac{3}{2},\frac{3}{2})$ resonance, which can
be regarded as the first member of a family of even parity resonances,
which correspond to the following sequence of values of the
angular momentum $L$ of the proton: $L = 0^+,2^+,4^+,(6^+,8^+)$.
Here we prefer to distinguish and keep separated the last two members
of the sequence, with $J^p=\frac{15}{2}^+,\frac{19}{2}^+$, in view
of the considerations which we are going to develop below. In fact, as the
phenomenological analysis of Hendry shows \cite{Hendry1}, the linear
rising of the Regge trajectory in the $\pi^+$-p elastic scattering,
is violated by the members of the family with high $J$, in particular,
with $J^p=\frac{15}{2}^+,\frac{19}{2}^+$. One of the purposes of the
present article is precisely to show that this phenomenon, in the
specific case of $\pi^+$-p elastic scattering, corresponds to a transition
from sharp resonances, which lie on a trajectory of standard form
$J=\alpha_0+\alpha' m^2$ ($\alpha'\sim 1/(\mathrm{GeV})^2$), to
surface waves. The resonances are properly described by Regge poles,
and can be associated with a vibrational spectrum generated by the
symplectic algebra $\fsp(3,\R)$.
Differently, surface waves are described by an other class of poles, which
we call Sommerfeld poles, whose location and motion in the complex angular momentum (CAM) plane
are radically different from those of the resonance poles.

The theory we present in Section \ref{se:theory} splits into two
parts: in the first part we show that the resonances (in particular,
the first three states with $J^p=\frac{3}{2}^+,\frac{7}{2}^+,\frac{11}{2}^+$)
lie on a linear trajectory which can be associated with a vibrational type
spectrum. To this end we develop a non--relativistic quark model,
studying the three--body dynamics generated by the quarks (\emph{uud}) confined by
harmonic--type potentials. We are thus led to the $\SU(3)$ classification of
three--particle states; in particular, by removing the degeneracies of the
harmonic oscillator, we obtain an Elliott--type rotational spectrum \cite{Elliott2}. But, it
is easy to see that a trajectory of the form $J=\alpha_0+\alpha' m^2$
is far from being generated by a rotational spectrum, in spite of the rule $\Delta J = 2$.
It rather appears to be closer related to a relativistic extension of a harmonic
oscillator model, where $L$ is proportional to the energy, i.e., $L\propto E$.
In fact, note that, accounting for the relativistic kinematics, one could expect
to get $L\propto E^2$, which yields a behavior which corresponds to a Regge--type trajectory \cite[pag. 91]{Collins}.
One is thus led to an enlargement of the spectrum generating group (SGG) $\SL(3,\R)$
toward precisely the SGG $\Sp(3,\R)$. Accordingly we obtain a vibrational--like spectrum,
associated with a symplectic group, which represents, \emph{at the non--relativistic level},
a model of a linear trajectory.

In the elastic $\pi^+$--p collision the first few unstable states are
sharp resonances (as a typical example keep in mind the $\Delta(\frac{3}{2},\frac{3}{2})$ resonance),
where only one partial wave is neatly dominant. But, as the energy increases,
inelastic and reaction channels open: the scenario changes drastically.
The elastic unitarity condition does not hold anymore, and the target
may be thought of as a ball totally or partially opaque at the center and with
a semitransparent shell at the border. At these energies the colliding beam undergoes diffraction.
The grazing trajectories hitting the target split into two rays:
one ray leaves the target tangentially, while the other one propagates along the
edge, creeping around the target. We have the surface wave phenomenon.
Instead of a single dominant partial wave, this phenomenon is due to a
packet of partial waves. Accordingly, the standard Fourier--Legendre
expansion in partial waves of the scattering amplitude converges slowly,
and it is therefore unsuited to describe surface waves.
A Watson--type resummation of the partial wave expansion becomes needed, and can be
performed as Sommerfeld did in connection with the diffraction of
radio waves around the earth \cite{Sommerfeld}. Following Sommerfeld,
surface waves can still be described in terms of poles, but these poles manifest
features and behavior rather different from those describing resonances, e.g., instead
of being located close to the real axis in the CAM--plane, they lie on a line
which is nearly parallel to the imaginary axis. Therefore, in order to distinguish
clearly these two different classes of poles, we call \emph{Regge poles} those
referring to resonances, and \emph{Sommerfeld poles} those referring to
surface waves. The analysis leading to the Sommerfeld poles is developed in detail
in Subsection \ref{subse:surface}.

In Section \ref{se:phenomenological} we show
that it is indeed the difference between these two classes of poles which
explains the phenomenological results. More
precisely, Section \ref{se:phenomenological} is split into two parts: in the first part
we study the first three resonances ($J^p=\frac{3}{2}^+,\frac{7}{2}^+,\frac{11}{2}^+$), and show
that they lie on a straight line trajectory of Regge type. The fits of
the cross--sections are analyzed by means of poles in the CAM--plane, and results in agreement
with the first part of the theory are then obtained. In the second part of Section \ref{se:phenomenological}
the cross--section is analyzed at higher energy, and we study in detail
the effects of the surface waves with particular attention to the transition
from resonances to creeping waves. Even in this case we obtain results which
agree with the second part of the theory.
Finally, in Section \ref{se:conclusions} some conclusions are summarized.

\section{The theory}
\label{se:theory}
\subsection{Spectrum of the resonances in $\boldsymbol{\pi^+}$--$\boldsymbol{\rm p}$ elastic scattering}
\label{subse:spectrum}

\subsubsection{Non--relativistic Schr\"odinger dynamics of three bodies confined by harmonic potentials}
\label{subsubse:nonrelativistic}

In the quark model the proton is composed by two quarks \emph{u} and one quark \emph{d}. The mass
of these quarks is approximately equal, and therefore we may treat, with good approximation,
these three bodies as having the same mass.
The possible mathematical tools for tackling the problem are:
the hyperspherical formalism \cite{Richard}, and the Faddeev equations \cite{Richard}.
Since the study of symmetries and, accordingly, the group theoretical methods, play
in our analysis a very relevant role, the hyperspherical formalism appears
more suitable to our purpose.
The hyperspherical method has been used frequently, and is well--known \cite{Richard}.
Therefore, we run through this argument very rapidly, and we will give
the necessary results in a form appropriate to our successive group theoretical analysis.

Let us consider three particles of equal mass $m$, whose positions are described by the vectors
$\br_k=(x_k, y_k, z_k)$, $(k=1,2,3)$. The kinetic energy operator reads
\beq
\label{1}
T=-\frac{1}{2m}(\Delta_1+\Delta_2+\Delta_3) \qquad (\hbar=1),
\eeq
where $\Delta_k=\partial^2/\partial x_k^2+\partial^2/\partial y_k^2+
\partial^2/\partial z_k^2$ $(k=1,2,3)$. We now introduce the Jacobi and center of mass
coordinates, defined as follows:
\begin{subequations}
\label{sub234}
\begin{eqnarray}
&&\bxi_1 = \frac{\br_1-\br_2}{\sqrt{2}}, \label{2} \\
&&\bxi_2 = \left(\frac{2}{3}\right)^{1/2}\left(\frac{\br_1+\br_2}{2}-\br_3\right), \label{3} \\
&&\bR_{\rm c.m.} = \frac{\br_1+\br_2+\br_3}{3} \qquad\qquad |\br_k|=\sqrt{x_k^2+y_k^2+z_k^2}. \label{4}
\end{eqnarray}
\end{subequations}
The kinetic energy operator can be written in these coordinates as
\beq
\label{5}
T=-\frac{1}{2m}\left(\Delta_{\bxi_1}+\Delta_{\bxi_2}+\frac{1}{3}\Delta_{\bR_{\rm c.m.}}\right),
\eeq
where
\begin{subequations}
\label{sub67}
\begin{eqnarray}
&&\Delta_{\bxi_i} \equiv \frac{\partial^2}{[\partial(\bxi_i)_x]^2}+
\frac{\partial^2}{[\partial(\bxi_i)_y]^2}+\
\frac{\partial^2}{[\partial(\bxi_i)_z]^2} \qquad (i=1,2), \label{6} \\
&&\Delta_{\bR_{\rm c.m.}} \equiv \frac{\partial^2}{[\partial(\bR_{\rm c.m.})_x]^2}+
\frac{\partial^2}{[\partial(\bR_{\rm c.m.})_y]^2}+
\frac{\partial^2}{[\partial(\bR_{\rm c.m.})_z]^2}, \label{7}
\end{eqnarray}
\end{subequations}
$(\bxi_i)_x, (\bxi_i)_y, (\bxi_i)_z$ and $(\bR_{\rm c.m.})_x,
(\bR_{\rm c.m.})_y, (\bR_{\rm c.m.})_z$ denoting the $x,y,z$
components of the vectors $\bxi_i$ and $\bR_{\rm c.m.}$,
respectively. Then, the kinetic energy of the center of mass can
be separated from that of the relative motion $T_R$:
\beq
\label{8}
T_R=-\frac{1}{2m}\left(\Delta_{\bxi_1}+\Delta_{\bxi_2}\right).
\eeq
Now, it is convenient to combine the vectors $\bxi_1$ and
$\bxi_2$ into a single vector $\bXi={\bxi_1 \choose \bxi_2}$, whose
Cartesian components will be denoted by
$\Xi_1,\Xi_2,\ldots,\Xi_6$. We can now consider a sphere
embedded in $\R^6$, whose squared radius is $\rho^2=\bxi_1^2+\bxi_2^2$
and, accordingly, represent the components of $\bXi$ in terms of
the spherical coordinates $(\rho,\theta_1,\ldots,\theta_5)$ as
follows:
\beq
\begin{split}
& \Xi_1 = \rho\sin\theta_5\sin\theta_4\cdots\sin\theta_1, \\
& \Xi_2 = \rho\sin\theta_5\sin\theta_4\cdots\cos\theta_1, \\
& \cdots \cdots \cdots \\
& \Xi_5 = \rho\sin\theta_5\cos\theta_4, \\
& \Xi_6 = \rho\cos\theta_5.
\end{split}
\label{9} 
\eeq
In terms of spherical coordinates the Laplace--Beltrami operator reads \cite{Vilenkin}
\beq
\begin{split}
& \Delta = \frac{1}{\rho^5}\frac{\partial}{\partial\rho}\left(\rho^5\frac{\partial}{\partial\rho}\right)
+\frac{1}{\rho^2\sin^4\theta_5}\frac{\partial}{\partial\theta_5}
\left(\sin^4\theta_5\frac{\partial}{\partial\theta_5}\right) \\
& \quad +\frac{1}{\rho^2\sin^2\theta_5\sin^3\theta_4}\frac{\partial}{\partial\theta_4}
\left(\sin^3\theta_4\frac{\partial}{\partial\theta_4}\right) \\
& \quad + \cdots +
\frac{1}{\rho^2\sin^2\theta_5\sin^2\theta_4\cdots\sin^2\theta_2}
\frac{\partial^2}{\partial\theta_1^2},
\end{split}
\label{10}
\eeq
and, by separating the radial part from the angular one, we have:
\beq
\label{11}
\Delta=\frac{1}{\rho^5}\frac{\partial}{\partial\rho}\left(\rho^5\frac{\partial}{\partial\rho}
\right) + \frac{1}{\rho^2}\Delta_0,
\eeq
where $\Delta_0$ is the Laplace--Beltrami operator acting on the unit
sphere $\SSS^5$ embedded in $\R^6$ \cite{Vilenkin}.\\
Let us now introduce the harmonic polynomials of degree $j$,
which may be written as $\rho^j\Theta_j(\theta_1,\ldots,\theta_5)$ \cite{Vilenkin}.
Then, from (\ref{11}) we obtain
\beq
\begin{split}
& \Delta\left[\rho^j\Theta_j(\theta_1,\ldots,\theta_5)\right] \\
&\quad=j(j+4)\rho^{(j-2)}\Theta_j(\theta_1,\ldots,\theta_5)
+\rho^{(j-2)}\Delta_0\Theta_j(\theta_1,\ldots,\theta_5)=0,
\end{split}
\label{12}
\eeq
which gives
\beq
\label{13}
\Delta_0\Theta_j(\theta_1,\ldots,\theta_5)=
-j(j+4)\Theta_j(\theta_1,\ldots,\theta_5).
\eeq
Next, we introduce a potential of the following form:
\beq
\label{14}
V(\rho)=G\left[|\br_1-\br_2|^2+|\br_1-\br_3|^2+|\br_2-\br_3|^2\right]=3G\rho^2,
\eeq
which is a \emph{confining potential of harmonic type}.

The Schr\"odinger equation reads:
\beq
\label{schro}
H\psi = (-\frac{1}{2m}\Delta + V) \psi = E \psi,
\eeq
where $E$ denotes the energy, and the operator $\Delta$ and the potential
$V$ are defined by formulae (\ref{10}) and (\ref{14}), respectively.
Next, by separating in the wavefunction $\psi(\rho;\theta_1,\ldots,\theta_5)$
the radial variable from the angular ones, we have the following equations:
\begin{eqnarray}
&&\frac{1}{\rho^5}\frac{d}{d\rho}\left(\rho^5\frac{dR_j}{d\rho}\right)
-\frac{j(j+4)}{\rho^2}R_j+2m[E-V(\rho)]R_j=0, \label{15} \\
&&\Delta_0\Theta_j(\theta_1,\ldots,\theta_5)=-j(j+4)\Theta_j(\theta_1,\ldots,\theta_5),
\label{16}
\end{eqnarray}
The solutions of Eq. (\ref{15})
are:
\begin{eqnarray}
R_j(\rho) &=& \rho^j\exp\left(-\frac{1}{2}\sigma^2\rho^2\right), \qquad\sigma=(mK)^{1/4},
\label{17} \\
E_j &=& (j+3)\omega, \qquad\omega=\sqrt{\frac{K}{m}}, \label{18}
\end{eqnarray}
where $K=6G$.\\

\begin{remark}
\label{rem:1}
It is worth noting that the zero--point energy in the harmonic spectrum in (\ref{18}) is:
$E_0=3\omega$ ($\hbar=1$), $\omega=\sqrt{K/m}$, $K=6G$,
$G=V(\rho)/(3\rho^2)$, and it is therefore strictly related to the strength of the
confining potential.
\end{remark}

Analogously to what done for the coordinates of the particles, we introduce for the momenta the vector
$\bGamma={\bp_{\bxi_1} \choose \bp_{\bxi_2}}$, where
\begin{subequations}
\label{19}
\begin{eqnarray}
&& \bp_{\bxi_1} = \frac{\bq_1-\bq_2}{\sqrt{2}}, \label{19a} \\
&& \bp_{\bxi_2} = \left(\frac{2}{3}\right)^{1/2}\left(\frac{\bq_1+\bq_2}{2}-\bq_3\right), \label{19b} \\
&& \bp_{\bR_{\rm c.m.}} = \frac{\bq_1+\bq_2+\bq_3}{3}, \label{19c}
\end{eqnarray}
\end{subequations}
and $\bq_k=m\dot{\br}_k$ ($\dot{\br}_k=d\br_k/dt;~k=1,2,3$) are the momenta of the particles.
Next, we define the \emph{grand--angular--momentum--tensor} \cite{Dragt}, which
is an antisymmetric $6\times 6$ tensor whose elements are
\beq
\label{20}
\Lambda_{kl}=\Xi_k\Gamma_l-\Xi_l\Gamma_k
\qquad (k,l=1,2,\ldots,6).
\eeq
Now, we associate to the momentum $\bGamma$ the following differential operators:
\beq
\label{21}
\Gamma_k=-\rmi\frac{\partial}{\partial\Xi_k} \qquad
(\hbar=1;k=1,2,\ldots,6).
\eeq
Then the following commutation rules hold true:
\beq
\label{22}
\left[\Xi_k,\Gamma_l\right]=\rmi\delta_{kl} \qquad
(k,l=1,2,\ldots,6).
\eeq
Substituting (\ref{21}) into (\ref{20}), we have
\beq
\label{23}
\Lambda_{kl}=-\rmi\left(\Xi_k\frac{\partial}{\partial\Xi_l}
-\Xi_l\frac{\partial}{\partial\Xi_k}\right) \qquad
(k,l=1,2,\ldots,6),
\eeq
and the following commutation rules \cite{Dragt}:
\begin{subequations}
\label{24}
\begin{eqnarray}
&& [\Lambda_{kl},\Lambda_{mn}]=0 \qquad k\neq l\neq m\neq n \label{24a} \\
&& [\Lambda_{kl},\Lambda_{lm}]=-\rmi\Lambda_{km} \label{24b} \\
&& \Lambda_{kl}=-\Lambda_{lk}. \label{24c}
\end{eqnarray}
\end{subequations}
We can now introduce the quantity
\beq
\label{25}
\Lambda^2=\frac{1}{2}\sum_{k,l=1}^6 \Lambda_{kl}^2,
\eeq
which satisfies the following commutation rules:
\beq
\label{26}
[\Lambda^2,\Lambda_{kl}]=0 \qquad (k,l=1,2,\ldots,6).
\eeq
Finally, rewriting $\Lambda^2$ in terms of spherical coordinates, and in view of (\ref{16}), we obtain
\beq
\label{27}
\Lambda^2\Theta_j=-\Delta_0\Theta_j=j(j+4)\Theta_j.
\eeq

\subsubsection{Permutation group on three objects and classification of three--particle states according to $\SU(3)$: rotational bands}
\label{subse:permutation}

The three--body motion may be described by means of the relative position
vector between the particles 1 and 2, and of the vector connecting the particle 3 with
the center of mass of the pair 1--2, i.e., using the Jacobi coordinates, if and only
if we can treat all three particles \emph{symmetrically}.

The group $\SSSS_3$ of permutations on three objects embraces six elements: the
three transpositions $P_{ik}$ (or interchange of particles $i$ and $k$, $i<k$),
and the three cyclic permutations: $C=P_{23}P_{12}$, which performs
the transformation $123 \rightarrow 312$, $C^2$ whose effect is $123 \rightarrow 231$,
and, finally, $C^3=e$, i.e., the identity $123 \rightarrow 123$. The transpositions $P_{ik}$
have matrix representation with determinant equal to $-1$, whereas the cyclic permutations
have determinant equal to $+1$. Therefore the
elements of $\SSSS_3$ may be split into two classes, depending on the
sign of their determinant.
Moreover, the cyclic permutations enjoy a continuous connection to the identity, which
lies entirely within the group $\SSSS_3$ \cite{Simonov}.

It is easy to see that the transformation of permutation takes us out of the space
of the components of $\bxi_1$ alone or of $\bxi_2$ alone, and mix the two sets of components \cite{Simonov}.
But $\rho^2=\bxi_1^2+\bxi_2^2$ is an invariant
under both three--dimensional rotations and permutations, and, in general, under all
six--dimensional rotations \cite{Simonov}. Then a remarkable
fact is that the elements of the permutation group yield to rotations in $\R^6$.
\emph{We are thus led to consider groups which act transitively on the sphere
$\SSS^5$ embedded in $\R^6$}. The first obvious choice in this direction is to consider the group
$\SO(6)$: $\SSS^5$ can be regarded as the space $\SO(6)/\SO(5)$ indeed.
Consider now the Lie algebra $\fso(6)$ associated with the group $\SO(6)$.
As well--known, the Lie algebra $\fg$ of the group $\GG$ can be identified with
the tangent space to $\GG$ at the identity element, i.e., $\fg \simeq T\GG|_e$.
In particular, $\fso(6)$ is the Lie algebra consisting of all
$6\times 6$ real skew--symmetric matrices, and is the real form of the
complex Lie algebra $D_3$ of dimension 15 (recall that the dimension of
$D_n$ is given by $n(2n-1)$ \cite{Barut}). But in Ref. \cite{Dragt} Dragt has proved that \emph{not all}
the elements of $\fso(6)$ treat all three particles equivalently. This
forces us to look for a subset of $\fso(6)$ whose elements satisfy the fundamental
requirement of treating all three particles equivalently. On the other hand,
exponentiating this subalgebra, we must obtain a subgroup of $\SO(6)$, which
acts transitively on $\SSS^5$. \emph{We have only one candidate which satisfies
this condition, the group $\SU(3)$}. In fact the space
$\C^n$ may be identified with the space $\R^{2n}$, by writing
out in a fixed order real and imaginary parts of the vector components in $\C^n$.
Therefore, $\SSS^5$ may also be regarded as the unit sphere
embedded in $\C^3$, and, accordingly, may be identified with the quotient
space $\SU(3)/\SU(2)$: $\SU(3)$ acts transitively on $\SSS^5$, indeed.
The Lie algebra $\fsu(3)$, associated with the group $\SU(3)$, consists
of all $3 \times 3$ skew--hermitian matrices $\cZ$ with $\Tr\cZ=0$.
It is one of the real forms of the complex Lie algebra $A_2$, whose dimension
is $8$ (recall that the dimension of $A_n$ is $n(n+2)$ \cite{Barut}).

\begin{remark}
\label{rem:2}
\rm
It is interesting to note that the Lie algebras $\fsu(3)$ and $\fsl(3,\R)$
are real forms of the same algebra $A_2$ \cite{Gilmore}. This fact is particularly
relevant in connection with the analysis which will be developed in the
next subsection.
\end{remark}

On the other hand Dragt \cite{Dragt} introduces a subset of $\fso(6)$ (i.e., a Lie
algebra which he denotes $L_1$) defined as the set of all elements $F \in \fso(6)$
which commute with the cyclic permutation $C=P_{23}P_{12}$, that is, satisfying
$[C,F]=0$. Dragt proves that all the elements of $L_1$ treat all three particles
with complete symmetry: if one permutes the particles, $L_1$ either remains unaffected
or undergoes a sign change. Even in this latter case it is still impossible to tell
which pair of particles has been interchanged, and which particle has been left alone \cite{Dragt}.
The algebra $L_1$ is nine--dimensional and isomorphic to the Lie
algebra associated with the $\rU(3)$ group. The operators of this algebra
may be expressed either in Cartesian form or in spherical tensor form.
If this latter form is used, then a scalar form can be separated from
the remaining components, and these latter components form a Lie algebra
$L_2$ of dimension $8$ isomorphic to the Lie algebra $\fsu(3)$.
\emph{We can thus conclude that the three--particle states can be completely classified by their
transformation properties according to the $\SU(3)$ group}.

This fact leads us to consider the sphere $\SSS^5$ as the unit sphere embedded
in $\C^3$, identified with the space $\SU(3)/\SU(2)$; accordingly, we
define the complex vectors
\begin{subequations}
\label{33}
\begin{eqnarray}
\bZ=\bxi_1+\rmi\bxi_2,\qquad\quad && \bZ^*=\bxi_1-\rmi\bxi_2, \label{33a} \\
\bPi=\bp_{\bxi_1}+\rmi \bp_{\bxi_2}, \qquad && \bPi^*=\bp_{\bxi_1}-\rmi \bp_{\bxi_2}.
\label{33b}
\end{eqnarray}
\end{subequations}
Then we have
\begin{subequations}
\begin{eqnarray}
\label{34}
&&\bZ\cdot\bZ^*=\bxi_1^2+\bxi_2^2=\rho^2, \label{34a} \\
&&\bPi\cdot\bPi^*=\bp_{\bxi_1}^2+\bp_{\bxi_2}^2=-\Delta. \label{34b}
\end{eqnarray}
\end{subequations}
Next, setting in Eqs. (\ref{17}) and (\ref{18}) $m=\hbar=1$,
$G=1/6$ ($K=1$), the total Hamiltonian can be written in the following form (see Eq. (\ref{schro})):
\beq
\label{35}
H=-\frac{1}{2}\Delta+V=\frac{1}{2}(\bPi\cdot\bPi^*+\bZ\cdot\bZ^*).
\eeq
In order to deal with the harmonic oscillator problem in the Fock space,
we introduce the vector creation and annihilation operators \cite{Dragt}
\begin{subequations}
\label{36}
\begin{eqnarray}
&&\bA^\dagger=\frac{1}{\sqrt{2}}(\bxi_1-\rmi\bp_{\bxi_1}),\qquad
\bA=\frac{1}{\sqrt{2}}(\bxi_1+\rmi\bp_{\bxi_1}),\label{36a} \\
&&\bB^\dagger=\frac{1}{\sqrt{2}}(\bxi_2-\rmi\bp_{\bxi_2}),\qquad
\bB=\frac{1}{\sqrt{2}}(\bxi_2+\rmi\bp_{\bxi_2}),\label{36b}
\end{eqnarray}
\end{subequations}
which satisfy the following commutation rules:
\begin{subequations}
\label{37}
\begin{eqnarray}
&&[A_k,A^\dagger_l]=\delta_{kl} \qquad (k,l=1,2,3), \label{37a} \\
&&[B_k,B^\dagger_l]=\delta_{kl}. \label{37b}
\end{eqnarray}
\end{subequations}
Therefore, the Hamiltonian may be written in terms of creation and annihilation
operators as
\beq
\label{38}
H=\frac{1}{2}(\bA^\dagger\cdot\bA+\bA\cdot\bA^\dagger+
\bB^\dagger\cdot\bB+\bB\cdot\bB^\dagger).
\eeq
Thanks to the commutation rules (\ref{37}) we have
\beq
\label{39}
\bA\cdot\bA^\dagger=3+\bA^\dagger\cdot\bA,\qquad\bB\cdot\bB^\dagger=3+\bB^\dagger\cdot\bB,
\eeq
and, therefore, we obtain
\beq
\label{40}
H=(\bA^\dagger\cdot\bA+\bB^\dagger\cdot\bB+3)=N_A+N_B+3,
\eeq
where $N_A$ and $N_B$ are the occupation numbers associated with the operators
$\bA^\dagger\cdot\bA$ and $\bB^\dagger\cdot\bB$, respectively.
Then, for the ground state $|0\rangle$, which is characterized by the conditions
$\bA|0\rangle=\bB|0\rangle=0$, we have $H|0\rangle=3|0\rangle$, which
represents the zero--point energy; correspondingly, the wavefunction is $\exp(-\rho^2/2)$.\\
Now, let $j_1$ and $j_2$ denote the eigenvalues of $N_A$ and $N_B$, respectively. Then from
Eqs. (\ref{18}) and  (\ref{40}) we have $j=j_1+j_2$.\\
For any group $G$ of linear transformations in a $n$--dimensional space, the tensors
of rank $r$ form a vector space of $n^r$ dimensions and constitute the basis for a
representation of the group $G$ \cite{Hamermesh}. By using permutation operators (Young
symmetrizers), this representation can be decomposed into irreducible representations of
$G$. Thus the whole space of the $r^\mathrm{th}$ rank tensors is reducible into subspaces
consisting of tensors of different symmetry. In the case of $\GL(3)$, the tableaux
for tensors of rank $r$ can contain at most three rows of length $f_1,f_2,f_3$
with $\sum f_i=r$, and $f_1\geqslant f_2\geqslant f_3 \geqslant 0$.
Consequently, an irreducible representation of $\GL(3)$ is characterized by
the partition $(f_1,f_2,f_3)$. Next, it can be shown that $(f_1,f_2,f_3)$
also serves as a label for irreducible representations of $\rU(3)$ \cite{Dragt}.\\
In general an irreducible representation of a group $G$, although being obviously
a representation of any subgroup $H$ of $G$, will not be irreducible with respect to $H$.
However, in the reduction of $\rU(n)$ to $\SU(n)$ (in particular of $\rU(3)$ to $\SU(3)$) the irreducible
representations of $\rU(n)$ remain irreducible under $\SU(n)$ \cite{Elliott2,Hamermesh}.
A simplification does occur nevertheless in that certain representations which were inequivalent under $\rU(n)$
become equivalent under $\SU(n)$ in view of the fact that $\SU(n)$ is a unimodular subgroup of $\rU(n)$ \cite{Elliott2}.
Consequently, for $\SU(3)$ the partition $(f_1,f_2,f_3)$ can be replaced by the differences: $k_1=f_1-f_3$,
$k_2=f_2-f_3$. Accordingly, the group $\SU(3)$ needs two rows to label its representations.
Putting $k_1=j_1+j_2$ and $k_2=j_1$ (i.e., $j_2=k_1-k_2$), we have the following Young pattern:
\begin{displaymath}
\yng(2,1)
\end{displaymath}
which corresponds to the representation $(j_1,j_2)$.
Coming back to the three-body dynamics, we now introduce the total angular momentum $\bL$ about the center
of mass \cite{Dragt}, i.e.,
\beq
\label{41}
\bL=\br_1\times\bq_1+\br_2\times\bq_2+\br_3\times\bq_3-\bR_{\rm c.m.}\times\bp_{\bR_{\rm c.m.}}.
\eeq
Note that $\bR_{\rm c.m.}$ and $\bp_{\bR_{\rm c.m.}}$ cannot vanish simultaneously since they do not commute.
In the center of mass we have $\bR_{\rm c.m.}=0$, while in the center of momentum frame
we have $\bp_{\bR_{\rm c.m.}}=0$. It follows from (\ref{41}) that $\bL$ may be interpreted either as the total angular momentum
about the center of mass or as the total angular momentum in the center of momentum frame. Finally,
it is important to remark that $\bL$ involves the three particles equivalently.
We are thus led to the following problem.

\begin{problem}
\label{prob:2}
Determine the $L$--values, $L(L+1)$ being the eigenvalues of $\bL^2$,
contained in the representation $(j_1,j_2)$ of $\SU(3)$.
\end{problem}

\begin{figure}
\begin{center}
\leavevmode
\includegraphics[width=12cm]{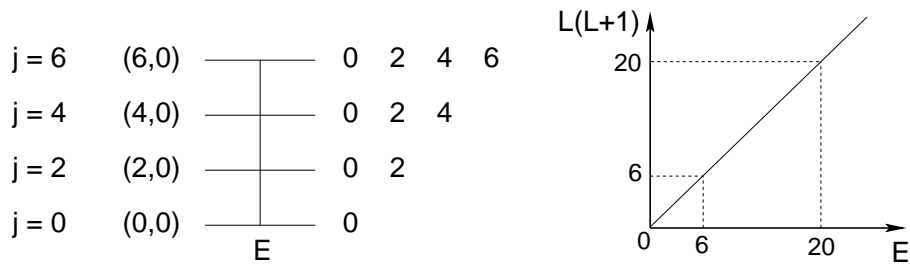}
\caption{\label{fig:1} Rotational--type spectrum, Elliott rule: $E\propto L(L+1)$.
}
\end{center}
\end{figure}

This problem can be rephrased as follows: determine what irreducible representations of the
group $\SO(3)$, which are labelled by $L$, occur in an irreducible representation of the
group $\SU(3)$. Weyl \cite{Elliott2,Weyl} has given a simple formula for the dimension of a representation of $\rU(n)$, i.e.,
\beq
\label{42}
\dim (f_1,f_2,\ldots,f_n)= \prod_{1\leqslant i<k\leqslant n}\left(\frac{f_i-f_k+k-i}{k-i}\right).
\eeq
Now, $(2L+1)$ is the dimension of the representation $D_L$ of the rotation group.
Then, Problem \ref{prob:2}  is solved by the equality
\beq
\label{43}
\prod_{1\leqslant i<k\leqslant 3}\left(\frac{f_i-f_k+k-i}{k-i}\right)=
\sum_{L}\mu_L (2L+1),
\eeq
where $\mu_L$ gives the number of times the representation $D_L$ occurs in a certain
representation of $\SU(3)$. Equality (\ref{43}) has been obtained by equating the characters
of the representations in the specific case of the unit element: recall, indeed, that the character
of a representation, corresponding to the unit element, gives the dimension of the representation.
In the present case we have $j_1=k_2=f_2-f_3$, and $j_2=k_1-k_2=f_1-f_2$. Therefore, from formula (\ref{43})
it follows
\beq
\label{44}
(j_1+1)(j_2+1)\left(\frac{j_1+j_2+2}{2}\right)=\sum_L \mu_L (2L+1).
\eeq
Since the l.h.s. of (\ref{44}) is symmetric in $j_1$ and $j_2$, it follows that $\dim (j_1,j_2)=\dim (j_2,j_1)$.
Now, we consider two cases:
\begin{itemize}
\item[(a)] Let $j_1=2n$ ($n$ integer) and $j_2=0$; then
\beq
\begin{split}
& \dim (j_1,j_2) = \dim (2n,0) \\
& \quad = (n+1)(2n+1)=\dim(D_0+D_2+\cdots D_{2n}).
\end{split}
\label{45}
\eeq
This means that the $L$--values that occur in the representation $(2n,0)$ are $L=0,2,4,\ldots,2n$
($\mu_L=1$). We have thus obtained a \emph{rotational} band of even parity.
\item[(b)] Let $j_1=2n+1$ ($n$ integer) and $j_2=0$; then
\beq
\begin{split}
& \dim (j_1,j_2) = \dim (2n+1,0) \\
& \quad = (n+1)(2n+3)=\dim(D_1+D_3+\cdots D_{2n+1}).
\label{46}
\end{split}
\eeq
This means that the $L$--values that occur in the representation $(2n+1,0)$ are $L=1,3,5,\ldots,2n+1$
($\mu_\ell=1$). We have thus obtained a \emph{rotational} band of odd parity.
\end{itemize}
Let us observe that levels with different values of $L$, but with the same value of $j=j_1+j_2$,
are degenerate.
In order to remove these degeneracies one can first construct the Casimir operator $C$ of the
$\SU(3)$ group. Recall that this operator is a scalar under the group. In our
case there are only two scalars, which one can form: one is the product $(\bL\cdot\bL)$
(where $\bL$ is the total angular momentum (\ref{41})); the second is the tensorial product
$(\bK\otimes\bK)$, where $\bK$ is a symmetric tensor which, in dyad notation, reads \cite{Dragt}:
\beq
\bK = \frac{1}{\sqrt{3}}\left[(\br_1-\br_2)\bq_3 + (\br_2-\br_3)\bq_1 + (\br_3-\br_1)\bq_2 + \mathrm{transpose}\right].
\label{N1}
\eeq
Note that $\bK$ involves the three particles equivalently \cite{Dragt}.
The Casimir operator is that combination of these two products which also
commutes with the group operators. Next one adds to the harmonic oscillator
Hamiltonian a term proportional to the Casimir operator $C$. This latter
will be diagonal in the $\SU(3)$ scheme with the same eigenvalue for all
states of a representation $(j_1,j_2)$. But, if these states are classified
by their angular momentum, then the energies for given $(j_1,j_2)$ follow the
rotational sequence $L(L+1)$: we obtain a rotational spectrum
(see Fig. \ref{fig:1}). We can thus say that the degeneracy is removed by a
splitting proportional to $L(L+1)$, as shown in Fig. \ref{fig:1}. But,
as remarked in the Introduction, the $J^p$ spectrum of even parity
in the $\pi^+$--p hadronic sequence shows a spectrum totally different
from that indicated in Fig. \ref{fig:1}.

It could be observed that the CAM model of resonances leads apparently
to the generation of rotational bands. In fact, a naive model of the resonance in the
CAM--plane can be roughly stated as follows: the incoming particle orbits
around the obstacle and if $2\pi R/\lambda=\ell$
($R=$ radius of the orbit, $\lambda=\mathrm{wavelength}=h/p$, $\ell$ integer $=$ angular momentum),
then, by taking the square, one obtains $E=\ell^2/(2I)$
($I$ being the moment of inertia), which yields approximately a rotational spectrum
(see Fig. \ref{fig:1}). Let us note that the simple geometrical condition
used above is equivalent to state that resonances occur at those discrete energies at which
the wavelength of the incoming particle is such that nodes of the wavefunction are put at the walls
of a well, whose radius $R$ is fixed; if this condition is satisfied, then nearly
stationary waves emerge. If the walls are not completely reflecting, the
lifetime of the resonance is finite. In fact, there is a wide phenomenological
evidence of rotational bands of resonances in non--relativistic ion collisions \cite{DeMicheli2},
where the CAM theory can be applied effectively, and a clear evidence
of Regge trajectories can be obtained \cite{DeMicheli1}. In these rotational bands
one can plot $L(L+1)$ versus $E$ (see Fig. \ref{fig:1}) and obtain a slope
given by $2I$ ($I$ is the moment of inertia). But, as we have already remarked,
this is not the case in the hadronic sequences, and specifically in the $\pi^+$--p
elastic resonances, where, in addition, the concept and theory of the relativistic
rotator seems to be far from a complete and satisfactory solution.

\subsubsection{Vibrational spectrum generated by the $\fsp(3,\R)$ algebra}
\label{subse:vibrational}

Let us come back to Remark \ref{rem:2} made in Subsection \ref{subse:permutation}: i.e., the Lie algebras $\fsl(3,\R)$
and $\fsu(3)$ are real forms of the same complex algebra $A_2$. In the previous subsection
we have seen the role played by $\fsu(3)$ in the three--body problem.
On the other hand, in the Introduction we have recalled that several authors \cite{Dothan,Sijacki}
suggested that the Lie algebra $\fsl(3,\R)$ should be considered in connection with the rule $\Delta J=2$
for the orbital angular excitations. In other words, it seems that both algebras play a relevant
role in the study of hadronic sequence spectrum. These arguments prompt to consider as
good candidate for describing the spectrum
of $\pi^+$--p elastic resonances the smallest Lie algebra containing both subalgebras
$\fsu(3)$ and $\fsl(3,\R)$, that is, the Lie algebra $\fsp(3,\R)$ associated with the
symplectic group $\Sp(3,\R)$. A $\Sp(3,\R)$ transformation can be regarded as a
\emph{general linear canonical transformation of both coordinates and momenta in phase space}.
We are thus led to look for a group of transformations of vectors in $\R^{2n}$ ($n=6$),
whose components are coordinates of positions and momenta.
Recalling once again that $\R^{2n}\equiv\C^n$, it is natural to explore the properties of the
group $\rU(6)$, acting on the unit sphere $\SSS^{2n-1}=\rU(n)/\rU(n-1)$ ($n=6$).
$\rU(n)$ is the group of matrices in $\GL(n,\C)$ (i.e., the group of the
complex $n\times n$ matrices with non--null determinant), which leave invariant the Hermitian
form: $x_1 \bar{y}_1+\cdots x_n \bar{y}_n$ in $\C^n$. Now, the field $\C$ can be replaced with the
quaternion field $\HH$ by doubling $\C$ (i.e., obtaining $\C^2$) to get $\HH$ (recall
that the algebras $\R$, $\C=\R^2$, $\HH=\C^2$ are all metric algebras). We are thus led to the group
$\rU^{\HH}(n)$ ($n=3$), which leaves invariant the Hermitian form
\beq
\label{47}
(\bxi,\bbeta)=\xi_1\bar{\eta}_1+\cdots+\xi_n\bar{\eta}_n,
\eeq
with
$\bxi=(\xi_1,\ldots,\xi_n)\in\HH^n,\bbeta=(\eta_1,\ldots,\eta_n)\in\HH^n, (n=3)$.
The group $\rU^\HH(n)$ may be regarded as a group of complex matrices, since any quaternion may be identified
with a pair $(u,v)$ of complex numbers ($\xi=u+v\rmj$). Let
\begin{subequations}
\label{49}
\begin{eqnarray}
\xi_1 &=& x_1+x_{n+1}\rmj,\,\cdots\,,\xi_n=x_n+x_{2n}\rmj, \label{49a} \\
\eta_1 &=& y_1+y_{n+1}\rmj,\,\cdots\,,\eta_n=y_n+y_{2n}\rmj, \label{49b}
\end{eqnarray}
\end{subequations}
then we have
\beq
\begin{split}
& \xi_1\bar{\eta}_1+\cdots+\xi_n\bar{\eta}_n =
[x_1\bar{y}_1+\cdots+x_n\bar{y}_n+x_{n+1}\bar{y}_{n+1}+\cdots+x_{2n}\bar{y}_{2n}] \\
& \quad +[(x_{n+1}y_1-x_1 y_{n+1})+\cdots+(x_{2n}y_n-x_n y_{2n})]\rmj,
\label{50}
\end{split}
\eeq
since $\overline{u+v\rmj}=\bar{u}-v\rmj$ and $v\rmj=\rmj\bar{v}$.
Therefore each element of the group $\rU^\HH(n)$ conserves the Hermitian form
$x_1\bar{y}_1+\cdots+x_{2n}\bar{y}_{2n}$ and the antisymmetric form
$(x_{n+1}y_1-x_1 y_{n+1})+\cdots+(x_{2n}y_n-x_n y_{2n})$. Reciprocally, if a matrix
is unitary and symplectic then, regarded as a transformation of $\HH^n$, conserves the form
$\xi_1\bar{\eta}_1+\cdots+\xi_n\bar{\eta}_n$ \cite{Postnikov}.
We can thus say that $\rU^\HH(n)$ is isomorphic to the group $\Sp(n,\C)\cap \rU(2n)\equiv \Sp(n)$
(in our case $n=3$, $\C^{2n}=\C^6=\HH^3$). Accordingly, the unit sphere $\SSS^{4n-1}$ ($n=3$)
may be identified with the space $\Sp(3)/\Sp(2)$. Next, taking the intersection
$\Sp(n)\cap \OO(2n)$, one obtains \cite{Postnikov}
\beq
\begin{split}
& \Sp(n) \cap \OO(2n) = \Sp(n,\C)\cap \rU(2n)\cap \OO(2n) \\
& \quad = \Sp(n,\R)\cap \rU(2n)\simeq \rU(n),
\end{split}
\label{53}
\eeq
which implies $\rU(n)\subset \Sp(n,\R)$. The following chain is particularly relevant in our analysis:
\beq
\label{54}
\Sp(3,\R)\supset \rU(3) \supset \SU(3) \supset \SO(3).
\eeq
In Refs. \cite{Rowe2,Park} a direct approach is proposed for obtaining the discrete spectrum associated
with $\Sp(3,\R)$, based on the fact that irreducible representations are
determined by their highest weight. This amounts to say that two irreducible representations with equal
highest weights are equivalent. Furthermore, the representation space is generated by the highest
weight vector by the action of the enveloping algebra. Now, the following propositions can be proved.

\begin{proposition}[Ref. \cite{Barut}]
\label{pro:4}
Every analytic irreducible representation of the real symplectic group $\Sp(n,\R)$ determines and is determined
by the highest weight $m=(m_1,m_2,\ldots,m_\nu)$, whose components are integers
satisfying the condition: $m_1\geqslant m_2\geqslant\cdots\geqslant m_\nu\geqslant 0$.
\end{proposition}

\begin{proposition}[Ref. \cite{Rowe2}]
\label{pro:5}
A highest weight state of $\fsp(3,\R)$ is a highest weight vector of its $\fu(3)$ subalgebra.
\end{proposition}

\begin{figure}[bt]
\begin{center}
\leavevmode
\includegraphics[width=11cm]{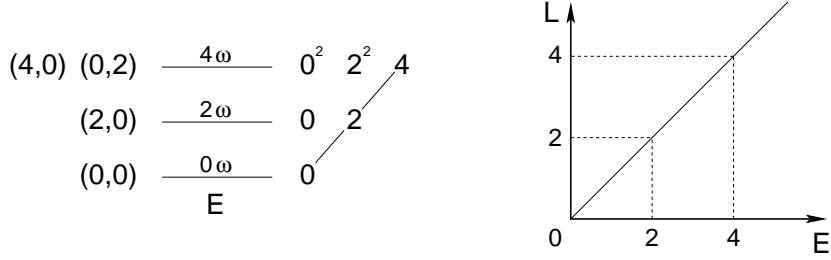}
\caption{\label{fig:2} Vibrational--type spectrum: $E\propto L$.}
\end{center}
\end{figure}

In Ref. \cite{Park} the infinitesimal generators of $\fsp(3,\R)$ are constructed by the use of
all the Hermitian quadratics in nucleon and momentum coordinates summed over particle index.
Next, in order to determine the irreducible unitary representation the authors pass to the
quadratics in the harmonic oscillator raising and lowering operators:
\begin{subequations}
\label{67}
\begin{eqnarray}
b_{ni}^\dagger &=& \left(\frac{m\omega}{2\hbar}\right)^{1/2}\left(x_{ni}-\frac{\rmi}{m\omega}p_{ni}\right), \label{67a}\\
b_{ni} &=& \left(\frac{m\omega}{2\hbar}\right)^{1/2}\left(x_{ni}+\frac{\rmi}{m\omega}p_{ni}\right). \label{67b}
\end{eqnarray}
\end{subequations}
Thus one obtains a basis of infinitesimal generators of $\fsp(3,\R)$:
\begin{subequations}
\label{68}
\begin{eqnarray}
A_{ij}&=&\sum_n b_{ni}^\dagger b_{nj}^\dagger, \label{68a} \\
C_{ij}&=&\frac{1}{2}\sum_n (b_{ni}^\dagger b_{nj} + b_{nj} b_{ni}^\dagger), \label{68b} \\
B_{ij}&=&\sum_n b_{ni} b_{nj}. \label{68c}
\end{eqnarray}
\end{subequations}
The $A_{ij}$ operators are $2\hbar\omega$ raising operators,
the $B_{ij}$ operators are $2\hbar\omega$ lowering operators, and
the $C_{ij}$ are $0\hbar\omega$ $\fu(3)$ operators.
Next, it is convenient to represent the $\fsu(3)$ content by the indexes $\lambda_0=N_1-N_2$,
$\mu_0=N_2-N_3$, where $\lambda_0$, $\mu_0$ are non negative integers with $N_1\geqslant N_2 \geqslant N_3$,
and $N_0=N_1+N_2+N_3$ is the harmonic oscillator eigenvalue.
Let $\cH_{N_0(\lambda_0,\mu_0)}$ denote the $\fsp(3,\R)$ representation space: it is necessary to decompose
this space into subspaces irreducible with respect to the unitary subalgebra $\fu(3)$.
Let $\cH^{(0)}_{N_0(\lambda_0,\mu_0)}$ denote the subspace containing the highest weight vector,
which transforms according to the $N_0(\lambda_0,\mu_0)$ irreducible representation of $\fu(3)$.
Since $A_{ij}$ is a $2\hbar\omega$ raising operator and $\cH_{N_0(\lambda_0,\mu_0)}$ is spanned by the polynomials in the
$A_{ij}$'s acting on the $N_0\hbar\omega$ oscillator space $\cH^{(0)}_{N_0(\lambda_0,\mu_0)}$, one concludes that
the only possible oscillator eigenvalues are $N_0\hbar\omega$, $(N_0+2)\hbar\omega$,
$(N_0+4)\hbar\omega, \ldots, (N_0+2r)\hbar\omega$ (see \cite[Theorem 2.2]{Rowe2}).
The further reduction of the $2r\hbar\omega$ eigenspace into irreps of $\fsu(3)$
is accomplished by first expressing the $r^{\rm th}$ degree polynomials in the $A_{ij}$
as $\fsu(3)$ irreducible tensor operators \cite{Rosensteel2} (see also Refs. \cite{Rowe4,Rowe3}).\\
Rosensteel and Rowe work out the problem using the Bargmann--Moshinsky approach \cite{Bargmann}.
In our present treatment, the $\SU(3)$ representations are labelled by the Cartan indexes $j_1=f_2-f_3=N_A$ and
$j_2=f_1-f_2=N_B$; $j=j_1+j_2=N_A+N_B$.
The Rosensteel--Rowe procedure can be adapted to the present treatment
which is close to the Cartan--Dragt presentation. First note that $j=N_A+N_B$ is the state of
highest weight in the sense of Cartan \cite{Dragt}. The Hamiltonian can be written in the following form:
$H=\frac{1}{2}(\bA^\dagger \bA+ \bA\bA^\dagger+\bB^\dagger \bB+\bB\bB^\dagger)$;
next, one adds the raising operators $\bA^\dagger \bA^\dagger$
and $\bB^\dagger \bB^\dagger$, and the lowering operators $\bA\bA$ and $\bB\bB$. Starting from the state $(0,0)$,
and then applying polynomials of raising operators $\bA^\dagger \bA^\dagger$
(or equivalently $\bB^\dagger \bB^\dagger$) one obtains a spectrum of the harmonic
oscillator in a discrete series of the form $(N_0+n)\hbar\omega$,
where $N_0$ is the smallest eigenvalue, and $n$ is an even non--negative integer: $n=0,2,4,\ldots$.
Let us recall that $\dim (j_1,j_2) = \dim (j_2,j_1)$.
In this way we obtain a vibrational spectrum (i.e., $L\propto E$), as depicted in Fig. \ref{fig:2}.

Since quarks in nucleons move at very nearly the speed of light,
we can push forward this non--relativistic model and
formulate a conjecture by observing
that in relativistic kinematics the connection between the energy $T$ and
the momentum $p$ is given by $T^2=p^2+m^2$ ($c=1$), which is the relativistic
analog of the relation $T=p^2/(2m)$. Then one might expect to have, at the
relativistic level, $L\propto E^2$ (see \cite[p. 91]{Collins}), which corresponds to the
actual behavior observed in $\pi^+$--p elastic scattering for the first few resonances
(see Subsection \ref{subse:resonances}).

\subsection{Resonances, echoes and surface waves: from Regge poles to Sommerfeld poles}
\label{subse:surface}

The sole possibility of exploring the internal structure of the proton and, in particular,
the vibrational spectrum associated with the $\Sp(3,\R)$ group, is to analyze the
effects of the interaction of a colliding particle acting as a probe, which hits
the proton regarded as a target. A good choice is the $\pi^+$--p interaction.
Of course this interaction is a two--body problem, and it should be treated with
appropriate coordinates in the ambient space $\R^3$. On the other hand, all the
results concerning the internal structure of the proton have been obtained by
the use of the Jacobi coordinates in an ambient space $\R^6$. Currently we do not know how to
transfer the kinematical and dynamical results obtained in a geometry embedded
in $\R^6$ to a geometry embedded in $\R^3$. We are then forced to follow a
phenomenological approach and, more specifically, to introduce all what occurs
for elaborating a scattering theory which allows us to use the main tools
necessary for fitting and interpreting the experimental data, as phase--shifts
and cross--sections. From the analysis of these data we can explore the proton
structure and, in particular, we can check if the resonances, which
eventually appear in the scattering process, correspond to the vibrational spectrum
derived in Subsection \ref{subse:vibrational}.

With this in mind, we start from the following integro--differential equation of Schr\"odinger type:
\beq
\label{k1}
(\Delta+V_D(\bR))\chi(\bR)+\int_{\R^3} V(\bR,\bR')\chi(\bR')\,d\bR' = E \chi(\bR),
\eeq
where $\Delta$ is the Laplace operator in $\R^3$, $\bR$ is the coordinate
of the relative motion between the two interacting particles, and $\chi(\bR)$ represents
the relative motion wavefunction. In (\ref{k1}) two types of potentials have
been introduced: a local one $V_D(\bR)$, and a non--local one $V(\bR,\bR')$, which
is here assumed to depend only on the lengths of the vectors $\bR$ and $\bR'$ and on
the angle between them; $E$, in the case of the scattering process, represents
the scattering relative kinetic energy of the two interacting particles in the
c.m.s.. Finally, the constant $\hbar$ and the reduced mass $\mu$ do not appear,
corresponding to the simple choice of units: $\hbar=2\mu=1$.

In this treatment we neglect the spin of the proton, the Coulomb
potential, and we limit ourselves to consider the non--relativistic scattering
of spinless non--identical particles. The need
to introduce a non--local potential, in addition to a local one, derives from the
fact that beside the phenomenon of resonances also \emph{echoes}
are present in the $\pi^+$--p elastic scattering: see, in particular, the
echo connected with the $\Delta(\frac{3}{2},\frac{3}{2})$ resonance
(see Subsection \ref{subse:resonances} and Fig. \ref{fig:4}).
If we describe the scattering process by
means of phase--shifts, then an upward crossing through $\pi/2$ of a phase--shift corresponds to
a resonance, whereas a downward crossing through $\pi/2$ corresponds
to an echo. In a neighborhood of an echo, instead of having a
\emph{time delay} (proportional to the lifetime of the resonance) we have a
\emph{time advance}. In Ref. \cite{DeMicheli2,DeMicheli3} we have developed a detailed analysis
of the resonance--echo process in connection with ion collision. In
the present situation the phenomenon is quite similar: when the pion and the
proton \emph{get in contact} and successively penetrate each other, the composing
quarks come into play. In view of the fermionic character of the quarks and of the
Pauli principle, the wavefunction must be antisymmetrized with respect to
all particle exchanges. Exchange forces emerge indeed, and these lead to non--local
potentials, in close analogy with the ion collision theory \cite{DeMicheli2}.
Unfortunately we do not know the precise form of the potentials $V_D(\bR)$ and
$V(\bR,\bR')$; therefore we assume only very general properties for these potentials,
which allow us to develop a scattering theory and, in particular, to perform a suitable
Watson--type resummation of the partial wave expansion, which naturally leads
to the introduction of the complex angular momentum (CAM) technique and, specifically,
of the Regge poles. In this connection we require that:
\begin{itemize}
\item[(a)] $V_D(\bR)$ and $V(\bR,\bR')$ are real--valued; $V(\bR,\bR')$ is a symmetric
function: $V(\bR,\bR') = V(\bR',\bR) = V^*(\bR,\bR')$. Accordingly, the Hamiltonian
$H = (-\Delta+\cU)$, where
\beq
(\cU\chi)(\bR)=V_D(\bR)\chi(\bR)+\int_{\R^3}V(\bR,\bR')\chi(\bR')\,d\bR' \nonumber
\label{hamiltonian}
\eeq
is a time--reversal invariant and formally Hermitian operator;
\item[(b)] $V(\bR,\bR')$ is a function only of $R=|\bR|,R'=|\bR'|$ and $\cos\gamma=(\bR\cdot\bR')/(RR')$.
Accordingly, the Hamiltonian $H$ is a rotationally invariant operator;
\item[(c)] $V_D(\bR)$ decreases exponentially for $|\bR|\rightarrow +\infty$;
\item[(d)] $V(\bR,\bR')$ is measurable in $\R^3\times\R^3$, and a constant $\alpha$
exists such that
\beq
\label{k2}
\hspace{-1cm}
\quad C^2 = \int_{\R^3}(1+R^2)\,e^{2\alpha R}\,d\bR\int_{\R^3}(1+R'^2)R'^2\,e^{2\alpha R'}V^2(\bR,\bR')\,d\bR' < \infty.
\eeq
\end{itemize}
If these conditions are satisfied the scattering amplitude $f(E,\theta)$ (where
$E$ is the relative energy, and $\theta$ is scattering angle in the c.m.s.) may be
expanded in partial waves (see Refs. \cite{DeMicheli2,DeMicheli3}):
\beq
\label{k3}
f(E,\theta)=\sum_{\ell=0}^\infty (2\ell+1) a_\ell(E) P_\ell(\cos\theta),
\eeq
where $\ell$ denotes the relative angular momentum ($\hbar=1$),
$P_\ell(\cos\theta)$ are the Legendre polynomials, and $a_\ell$ reads:
\beq
\label{k4}
a_\ell(E)=\frac{e^{2\rmi\delta_\ell}-1}{2\rmi k},
\eeq
$\delta_\ell$ being the phase--shifts, and $k$ the momentum in the c.m.s..

\begin{figure}[t]
\begin{center}
\leavevmode
\includegraphics[width=8cm]{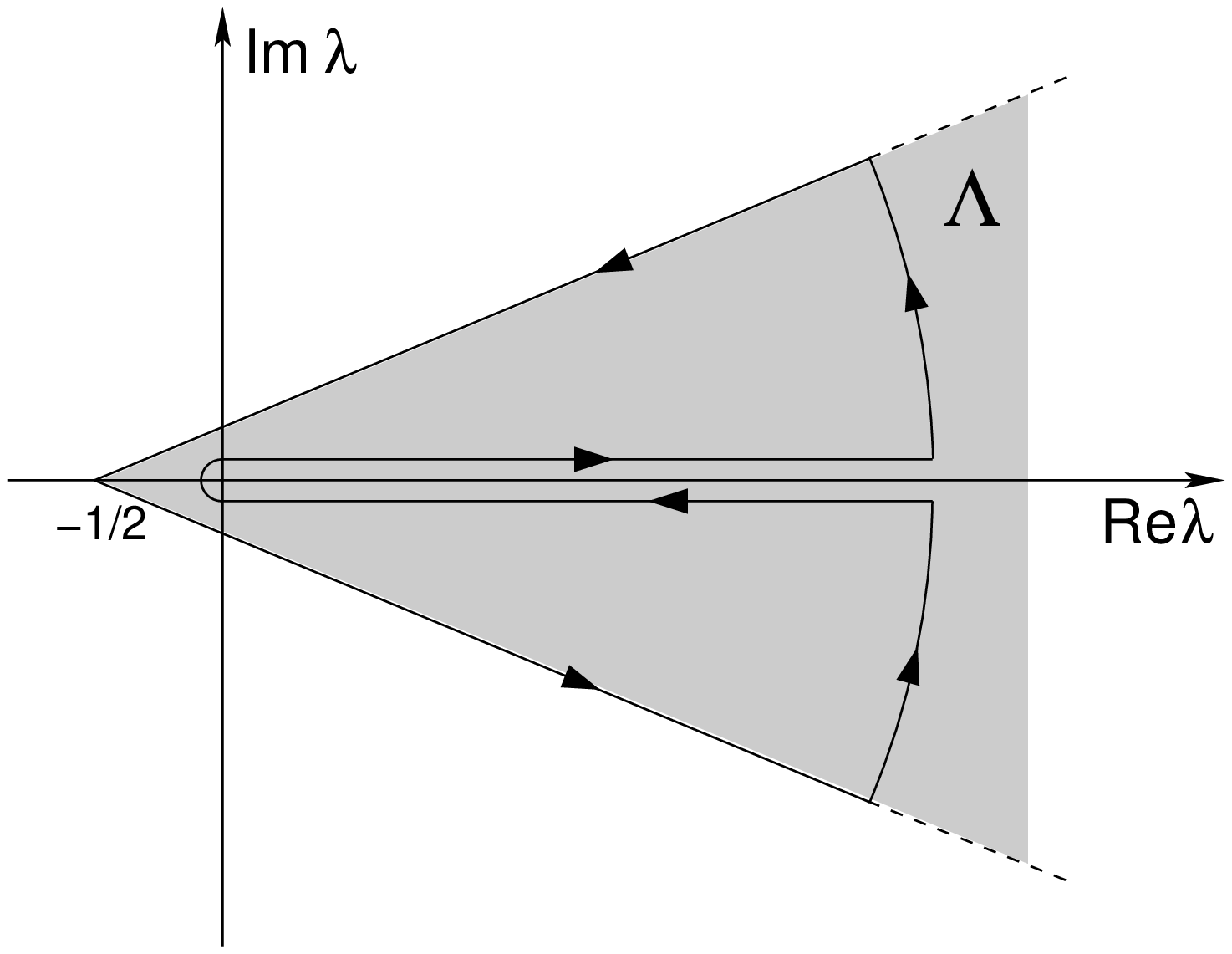}
\caption{\label{fig:3} The angular sector $\Lambda$ and the
integration path for the computation of the background integral.
\vspace{0.5cm}
}
\end{center}
\end{figure}

Much more delicate is the question concerning the Watson resummation of the partial wave expansion (\ref{k3}).
This resummation requires some rather restrictive conditions on the partial scattering amplitudes which, for instance,
must admit a unique interpolation in the CAM--plane in the sense of Carlson's theorem \cite{DeAlfaro}.
It has been proved that these constraints are satisfied by a rather limited class of potentials, notably by the
Yukawian class \cite{DeAlfaro}. It is well--known that this type of resummation splits the scattering amplitude
into two terms: a sum over poles of the scattering amplitude in the CAM--plane and a background integral, whose integration
path is usually taken to be parallel to the imaginary axis, i.e., from $-\frac{1}{2}-\rmi\infty$
to $-\frac{1}{2}+\rmi\infty$. We note, however, that in the present situation we are
working in the physical region of $\cos\theta$ (i.e., $-1\leqslant\cos\theta\leqslant 1$), and we are not interested
in the asymptotic behavior of the scattering amplitude for large transmitted momentum. Consequently,
in the background integral, we are not forced to take a path running along an axis parallel to the
imaginary axis of the CAM--plane, but we can rather close the Watson integration path along
the border of an appropriate angular sector $\Lambda$ in the CAM--plane (see Fig. \ref{fig:3}).
In spite of this relevant advantage we still need the potentials
in question to satisfy some additional conditions. In particular, if we expand the non--local potentials
$V(\bR,\bR')$ in series as follows:
\beq
\label{k5}
V(\bR,\bR')=\frac{1}{4\pi RR'}\sum_{\ell=0}^\infty(2\ell+1)V_\ell(R,R')P_\ell(\cos\gamma),
\eeq
($\cos\gamma=(\bR\cdot\bR')/(RR')$), then we have the so--called
\emph{partial potentials} $V_\ell(R,R')$, which are given by:
\beq
\label{k6}
V_\ell(R,R')=2\pi RR'\int_{-1}^1 V(R,R',\cos\gamma)P_\ell(\cos\gamma)\,d(\cos\gamma).
\eeq

Then, we must require the set of these \emph{partial potentials} $\{V_\ell\}_{\ell=0}^\infty$
to admit a unique Carlsonian interpolation $V(\lambda;R,R')$ ($\lambda$ denoting the complex angular momentum)
in the half--plane $\C^+_{(-1/2)} = \{\lambda\in\C : \Real\lambda > -\frac{1}{2}\}$.
Moreover, this interpolation should possess an exponential
decrease, for large values of $|\lambda|$, of the following form:
$V(\lambda;R,R')\sim e^{-\eta\Real\lambda}$ ($\eta>0$). If these conditions are satisfied\footnote{A
complete analysis of the conditions needed to perform a Watson--type resummation of the expansion of the
partial waves generated by non--local potentials along the lines indicated here, requires very long and
detailed mathematical proofs. This work has been published in a
mathematical physics journal \protect\cite{Bros}.}, then we can sum the partial wave expansion following the
method of Watson, and therefore transform the series over discrete values of $\ell$ into an integral encircling
the real positive semi--axis of the CAM--plane. Next, we can deform this integration path into a path composed
by arcs of circles and two straight lines which delimit an angular sector $\Lambda$ in the CAM--plane \cite{Bros}
(see Fig. \ref{fig:3}). The contributions along the arcs of circles can be proved to vanish, and then one remains
with a sum over poles and a background integral, whose path is composed by the two straight lines only.

We now lay great stress on the following point: \emph{while in the case of local
potentials (notably Yukawian) the poles are all located in the first quadrant of the CAM--plane \cite{DeAlfaro},
in the case of non--local potentials the poles can lie in both the first and the fourth quadrant
of the CAM--plane}. The standard interpretation of the poles lying in the first quadrant associates
these singularities to bound states if they are located on the real axis, and to resonances if they
lie inside the first quadrant. We can now associate the poles located in the fourth quadrant to the
\emph{echoes}. Let us indeed note that these poles are present only in the case of non--local potentials,
which are generated by those repulsive forces which emerge when the interacting particles, after getting
in contact, penetrate each other. The CAM polology allows us to describe both resonances and echoes
by pole singularities: the resonances by poles of the scattering amplitude which lie in the first quadrant; the echoes by poles lying
in the fourth quadrant \cite{DeMicheli3}. The imaginary part of the location of the poles lying in the first quadrant is positive
(i.e., $\Imag\lambda>0$) and corresponds to a \emph{time delay}; conversely, the imaginary part of the
location of the poles lying in the fourth quadrant is negative (i.e., $\Imag\lambda<0$),
and corresponds to a \emph{time advance}. Let us finally observe that in the classical Breit--Wigner
theory the echoes are described by the scattering by an impenetrable sphere. But the
roughness of this Breit--Wigner model can be easily verified just observing the behavior of the phase--shifts
in several scattering process: for instance, in the elastic $\alpha$--$\alpha$ scattering. In this case the
phase--shift $\delta_{(\ell=2)}$, crossing downward $\pi/2$ in correspondence of an echo, presents a clearly observable
concave behavior \cite{DeMicheli3}, instead of a linear one, as prescribed, at sufficiently high energy, by the asymptotic form
of the scattering by an impenetrable sphere.

Let us observe that also the type of Watson transformation which we have illustrated above leads to
split the total scattering amplitude into a sum over poles plus a background integral. The integration
path of this latter term is given by the straight lines which delimit the angular sector $\Lambda$
in the CAM--plane (see Fig. \ref{fig:3}). The poles can lie either in the first or in the fourth
quadrant. In the neighborhood of a resonance one pole, lying in the first quadrant, is dominant, and therefore it
is worthwhile to approximate the total scattering amplitude, in the energy domain around a resonance,
with the following formula (see also Refs. \cite{DeMicheli1,DeMicheli3}):
\beq
\label{k7}
f(E,\theta) \simeq g_r(E)\,\frac{P_{\lambda_r}(-\cos\theta)}{\sin\pi\lambda_r} \hsps (0<\theta\leqslant\pi),
\eeq
where $\lambda_r$ gives the location of the pole in the first quadrant (the subscript '$r$' is for recalling
that we refer to a resonance), and $P_{\lambda_r}$ is the Legendre function of the first kind.
This approximation fails forward (i.e., at $\theta=0$ and in a neighborhood of this angle), in view of the
cut $[1,+\infty)$ in the complex $\cos\theta$--plane carried by the function $P_{\lambda_r}(-\cos\theta)$ \cite{Sommerfeld}.
We see from formula (\ref{k7}) that if $|\Imag\lambda_r|\ll 1$, the amplitude $f(E,\theta)$
presents a pole--type behavior whenever $\Real\lambda_r$ crosses an integer. Accordingly,
a sharp resonance is observed in the cross--section. Amplitude (\ref{k7}) can be projected
on the $\ell^\mathrm{th}$ partial wave by means of the following formula \cite{Erdelyi}:
\beq
\label{Q2}
\hspace{-0.5cm}
\int_{-1}^{+1}P_\ell(z) P_{\lambda_r}(-z)\,dz = \frac{2\sin\pi\lambda_r}{\pi(\lambda_r-\ell)(\lambda_r+\ell+1)}
\hsps (\ell=0,1,2\ldots;\lambda_r\in\C),
\eeq
which holds true since $P_{\lambda_r}(-z)$ presents a  singularity of logarithmic type for $z=1$ \cite{Sommerfeld}.
We thus obtain the following expression for the partial wave scattering amplitude $a_\ell$:
\beq
\label{Q3}
a_\ell= \frac{e^{2\rmi\delta_\ell}-1}{2\rmi k}=
\frac{g_r}{\pi}\frac{1}{(\alpha_r+\rmi\beta_r-\ell)(\alpha_r+\rmi\beta_r+\ell+1)},
\eeq
where $\delta_\ell$ denotes the phase--shift, and $\lambda_r=\alpha_r+\rmi\beta_r$.
Next, whenever the elastic unitarity condition can be applied, we have the following
relationship among $g_r(E),\alpha_r,\beta_r$:
\beq
\label{Q4}
g_r(E)=-\frac{\pi}{k}\,\beta_r(2\alpha_r+1) \hsps (k^2=E=\mathrm{energy}),
\eeq
which, finally, yields
\beq
\label{Q5}
\delta_\ell=\sin^{-1}\frac{\beta_r(2\alpha_r+1)}
{\left\{\left[\left(\ell-\alpha_r\right)^2+\beta_r^2\right]\left[\left(\ell+\alpha_r+1\right)^2+\beta_r^2\right]\right\}^{1/2}}.
\eeq
When $\alpha_r(E)$ equals an integer and $\beta_r(E)$ is very small we have $\sin\delta_\ell\simeq 1$,
i.e., a resonance. Furthermore, formula (\ref{Q5}) can describe a sequence of resonances in various
partial waves. Indeed, the pole moves in the CAM--plane as a function of $E$, and resonances
occur whenever $\alpha_r(E)=\ell$, $\ell$ integer being the physical angular momentum ($|\beta_r|\ll 1$).
We have thus a trajectory which connects several resonances, and, moreover, \emph{the behavior of
this trajectory is closely connected with the symmetry properties of the interacting system}, as
we have shown in Subsection \ref{subse:spectrum}. These poles start close to the real positive semi--axis of the CAM--plane
(i.e., $|\Imag\lambda|\ll 1$), and constitute the first class of singularities, which correspond to
the well--known Regge poles. As the energy $E$ increases, $\Imag\lambda$ increases too in agreement
with the increasing values of the widths of the resonances. Accordingly, the distance of the poles from
the real axis of the CAM--plane becomes larger.

In the neighborhood of an echo we can proceed in a way very close to that
followed for describing the resonances. Then we write:
\beq
\label{kk1}
f(E,\theta) \simeq g_e(E)\,\frac{P_{\lambda_e}(-\cos\theta)}{\sin\pi\lambda_e} \hsps (0<\theta\leqslant\pi),
\eeq
where the subscript '$e$' stands for recalling that we refer to echoes.
Next, projecting the amplitude (\ref{kk1}) on the $\ell^{\rm th}$ partial wave, we obtain
\beq
\label{kk2}
a_\ell= \frac{e^{2\rmi\delta_\ell}-1}{2\rmi k}=
\frac{g_e}{\pi}\frac{1}{(\alpha_e-\rmi\beta_e-\ell)(\alpha_e-\rmi\beta_e+\ell+1)},
\eeq
where $\lambda_e=\alpha_e-\rmi\beta_e$ ($\beta_e>0$). Next, whenever the elastic unitarity condition can be applied,
use can be made of the following relationship among $g_e$, $\alpha_e$, $\beta_e$:
\beq
\label{kk3}
g_e(E)=\frac{\pi}{k}\,\beta_e(2\alpha_e+1) \hsps (k^2=E=\mathrm{energy}),
\eeq
which finally yields
\beq
\label{kk4}
\delta_\ell=\sin^{-1}\frac{-\beta_e(2\alpha_e+1)}
{\left\{\left[\left(\ell-\alpha_e\right)^2+\beta_e^2\right]
\left[\left(\ell+\alpha_e+1\right)^2+\beta_e^2\right]\right\}^{1/2}}.
\eeq
Let us note that, whenever $\alpha_e$ equals an integer, and $\beta_e$ is sufficiently small, then
$\delta_\ell\simeq -\pi/2$. Furthermore, let us remark that resonances and echoes take place
at different energies: a resonance occurs at an energy smaller than that of the corresponding echo.
Therefore, we may simply sum the two terms representing the resonances and the echoes respectively, and
neglect the interference effects. Indeed, at those values of energy where the resonance pole is dominant the
term due to the echo is negligible, and vice versa. In conclusion, we can write
\beq
\begin{split}
& \delta_\ell \simeq \sin^{-1}\frac{\beta_r(2\alpha_r+1)}
{\left\{\left[\left(\ell-\alpha_r\right)^2+\beta_r^2\right]\left[\left(\ell+\alpha_r+1\right)^2+\beta_r^2\right]\right\}^{1/2}} \\
& \quad +\sin^{-1}\frac{-\beta_e(2\alpha_e+1)}
{\left\{\left[\left(\ell-\alpha_e\right)^2+\beta_e^2\right]
\left[\left(\ell+\alpha_e+1\right)^2+\beta_e^2\right]\right\}^{1/2}}.
\end{split}
\label{kk5}
\eeq
We can therefore see that both resonances and echoes are described by pole singularities (in the first and in the
fourth quadrant of the CAM--plane, respectively), which act at different values of energy, in such a way that their effects can be separated
in a rather neat way. Finally, we remark that when the energy increases, the value of $\beta_e$ increases too;
accordingly, approaching the semiclassical limit, the effects of the echoes disappear in agreement with
the fact that they are typical \emph{quantum effects}.

As we already said several times, when the energy increases inelastic and reaction channels open, and the target
appears as a ball opaque at the center and semitransparent at the border.
We have reached the semiclassical limit, and consequently the effects of the echoes can be completely
neglected.
In order to understand the physical processes connected with the second class of poles, it is convenient and simpler
to start from the situation corresponding to the blackbody limit, when the target can
be regarded as a totally absorbing opaque ball. In this case the elastic scattering is mainly
due to diffraction undergone by the grazing rays. These latter, hitting the target, split
into two components: one ray leaves tangentially the diffracting body, and it is called
\emph{diffracted ray}, whereas the other ray describes a geodesic along the border of the ball.
The diffracted rays are strongly focused on the horizontal axis, which is a symmetry axis of the body and coincides
with the direction of the colliding beam, it is the axial caustic. For what concerns
the rays bending the target, we must distinguish between the ones travelling in counterclockwise
sense from those travelling in clockwise direction. The scattering angle
$\theta$ must be related to the surface angles $\theta_m^{(S^+)}$ and $\theta_m^{(S^-)}$, where
$\theta_m^{(S^+)}$ refers to the rays winding $m$ times around the target in
counterclockwise sense, and $\theta_m^{(S^-)}$ refers to the rays winding $m$ times around the
diffracting ball in the clockwise direction. We have:
\begin{subequations}
\label{Q6}
\begin{eqnarray}
\theta_m^{(S^+)} &=& \theta+2\pi m \qquad (m=0,1,2,\ldots), \label{Q6a} \\
\theta_m^{(S^-)} &=& 2\pi-\theta+2\pi m \qquad (m=0,1,2,\ldots). \label{Q6b}
\end{eqnarray}
\end{subequations}
Now, returning to formula (\ref{k7}), we first substitute to $P_{\lambda_r}(-\cos\theta)$
its asymptotic behavior, which holds for large values of $|\lambda|$,
i.e., we write (omitting for a while the subscript 'r') \cite{Erdelyi}:
\beq
\label{Q7}
P_\lambda(-\cos\theta)\simeq
\frac{e^{-\rmi[(\lambda+1/2)(\pi-\theta)-\pi/4]}+e^{\rmi[(\lambda+1/2)(\pi-\theta)-\pi/4]}}
{[2\pi(\lambda+1/2)\sin\theta]^{1/2}} \hsps (0<\theta<\pi).
\eeq
Next, we set $\lambda+\frac{1}{2}=\nu_1$ (the meaning of the subscript '1' in $\nu_1$ will
be clarified below); then we use the following expansion:
\beq
\label{Q8}
-\frac{1}{\sin\pi\lambda}=\frac{1}{\cos\pi\nu_1}=
2 e^{\rmi\pi\nu_1}\sum_{m=0}^\infty (-1)^m \,e^{\rmi 2m\pi\nu_1}
\hsps (\Imag\nu_1>0).
\eeq
Finally, from formulae (\ref{k7}), (\ref{Q7}), and (\ref{Q8}) we obtain, for $0<\theta<\pi$:
\beq
\label{Q9}
f(E,\theta)\simeq -\rmi G(E)\sum_{m=0}^\infty (-1)^m
\left[\frac{e^{\rmi\nu_1\theta_m^{(S^+)}}}{\sqrt{\left|\sin\theta_m^{(S^+)}\right|}}
-\rmi\frac{e^{\rmi\nu_1\theta_m^{(S^-)}}}{\sqrt{\left|\sin\theta_m^{(S^-)}\right|}}\right],
\eeq
where $G(E)=2g(E)e^{-\rmi\pi/4}/\sqrt{2\pi\nu_1}$.
The factor $(-1)^m$ in (\ref{Q9}) is due to the fact that both the counterclockwise and the clockwise
rays cross twice the axial caustic at each tour around the diffracting body; since
at each crossing there is a phase--shift of $e^{-\rmi\pi/2}$ for the counterclockwise rays and of
$e^{\rmi\pi/2}$ for the clockwise ones, we have precisely the factor $e^{\pm\rmi\pi}=-1$.
Furthermore, there is an additional phase--shift between the rays (counterclockwise and clockwise)
which explains the factor $-\rmi=e^{-\rmi\pi/2}$ for the counterclockwise rays and the
factor $\rmi^2=-1$ for the clockwise ones (see Ref. \cite{DeMicheli5} for a detailed mathematical
analysis). The term $\nu_1$ is given by: $\nu_1=\lambda+\frac{1}{2}=R(k+\rmi\gamma_1)$,
where $R$ is the radius of the diffracting body;
accordingly, we have $kR=\Real\lambda+\frac{1}{2}=\ell+\frac{1}{2}$, in agreement with the
semiclassical expression of the angular momentum.
The factor $\gamma_1$ is due to the damping of the flux of rays travelling along the
border of the target, and depends on the curvature of the diffracting body;
therefore, it is constant for a spherical ball. The damping is produced by the splitting
in two components of each ray bending the target at each point of the target (see also
Keller's geometrical theory of diffraction \cite{Hansen}).

Following Sommerfeld \cite{Sommerfeld}, the diffraction problem can be treated by starting from
Helmholtz's equation and looking for a solution which is continuous throughout the
exterior of a given bounded surface $\Sigma$, assuming arbitrarily prescribed boundary
values on $\Sigma$, and a suitable radiation condition at infinity \cite{Sommerfeld}.
In the case of a sphere of radius $R$ one can separate the variables, the angular
part of the solution is expressed in terms of Legendre polynomials, while the radial part is represented
by the Hankel functions of the first kind: $H_{n+1/2}^{(1)}(kr)$ ($k$ is the wavenumber, $r$ is
the distance from the center of the sphere, and $n$ is an integer). Sommerfeld imposes
a boundary condition of Dirichlet type on the surface of the sphere, i.e.,
\beq
\label{Q10}
H_{n+1/2}^{(1)}(kR)=0.
\eeq
The roots of (\ref{Q10}) \emph{lie in the positive imaginary $n$--half--plane, and are infinite in number} \cite{Sommerfeld}.
To emphasize that the index of these functions acquires complex values, we replace $n+\frac{1}{2}$
with $\nu$, and, accordingly, Eq. (\ref{Q10}) will be written $H_\nu^{(1)}(kR)=0$.
The roots of the equation $H_\nu^{(1)}(kR)=0$ are given, for $kR\gg 1$,
by the well--known formula obtained by van der Pol and Bremmer with the aid of the
Debye expansion for the Bessel function \cite{Keller2}:
\beq
\label{Q11}
\nu_m \simeq kR + 6^{-1/3} e^{\rmi\pi/3} (kR)^{1/3} q_m \quad
(kR \gg m=1,2,3,\ldots),
\eeq
where $q_m$ is the $m^\mathrm{th}$ zero of the Airy function $\mathrm{Ai}(q)$.
They are located close to a curve which tends to become parallel to the imaginary axis
of the $\nu$--plane. In particular, the root of (\ref{Q10}) which is closest to the real axis of the
$\nu$--plane corresponds to the value of $m=1$, and from (\ref{Q11}) we have in a first rough
approximation $\Real(\nu_1)\simeq kR$. In this approach the solution of the
diffraction problem can be written in terms of series which, however, converge very slowly.
Then Sommerfeld applies a Watson transformation to these series, transforming a sum over $n$
($n$ integer) into an integral along a suitable path in the complex $\nu$--plane.
The poles of this integral in the $\nu$--plane are precisely the roots of the
equation $H_\nu^{(1)}(kR)=0$.
The sum over residues at the poles located in the first quadrant of the $\nu$--plane is rapidly
convergent for values of the angle sufficiently large (i.e., backwards). In particular,
the term corresponding to the pole closest to the real axis is the dominant one, and it is, in general,
sufficient for describing the diffraction in the backward angular region.
\emph{This approximation is close to the one we derived from formulae (\ref{Q7}) and (\ref{Q9})},
and we return on this point with more details in Subsection \ref{subse:surfacewaves}.
In particular, the expression of $\nu_1$, used in formulae (\ref{Q8}) and (\ref{Q9}), is given by:
$\nu_1=\lambda+\frac{1}{2}=R(k+\rmi\gamma_1)$, and it can be regarded as a crude approximation
of the index $\nu_1$ given by formula (\ref{Q11}). Let us finally note that the subscript
'1' in the terms $\nu_1$ and $\gamma_1$ indicates that we are considering only
the pole closest to the real axis.

Now, we return to formula (\ref{k7}), which will be written in the following form:
\beq
f(E,\cos\theta) \simeq C(E)\,P_\lambda(-\cos\theta) \qquad (0<\theta\leqslant\pi),
\label{KK}
\eeq
observing that the \emph{elastic unitarity condition cannot be applied} in this context
and, moreover, the contribution of $|\sin\pi\lambda|^{-1}$, when $\Real\lambda$ crosses an integral value,
is strongly damped by a factor of the form $\exp(-\pi|\Imag\lambda|)$ where, at the blackbody
limit, $|\Imag\lambda|$ is of the order of $1$.
Next we note that the amplitude (\ref{KK}) is indeed factorized
into two terms: the first one, $C(E)$, gives the amplitude at $\theta=\pi$ as a
function of $E$ since $P_\lambda(1)=1$; the second factor describes the
backward angular distribution at fixed $E$. It follows that two types of
phenomenological fits are possible: at fixed angle $\theta=\pi$, and at
fixed energy (see Subsection \ref{subse:surfacewaves}). Let us however note that
the scheme of a totally opaque sphere is inadequate at least for energy ranges below the blackbody limit.
We must modify our model assuming that the interaction region presents an absorbing core
surrounded by a nearly transparent shell. Therefore, it must be taken into account
the contribution to the scattering amplitude also of those grazing rays that undergo
limiting refractions, and emerge after taking one or more shortcuts. To this purpose it
is worthwhile to mention the important work of Nussenzveig \cite{Nussenzveig}, who
studied the scattering by a transparent sphere. He uses a Debye expansion, which is
a representation of the scattering problem in terms of surface interactions, i.e.,
the scattering amplitude is decomposed into an infinite series of terms representing
the effects of successive internal reflections. In this way Nussenzveig has been able, in particular,
to explain the meteorological glory, giving a mathematical and numerical proof
of Van De Hulst's conjecture that surface waves are responsible for the
meteorological glory: specifically, diffracted rays having taken two shortcuts across the sphere.
In a geometrical optics approximation the Debye
expansion corresponds to the ray--tracing procedure;
using this method, the contributions due to the grazing rays which take $0,1,2,\ldots,n$ shortcuts
are summed up and, after retaining only the main contribution, the amplitude at backward angles can be written as \cite{Viano}
\beq
\label{Q12}
f(E,\theta) \simeq \sum_{p=0}^n C^{(p)}(E) P_\lambda(-\cos\theta).
\eeq
Since $P_\lambda(1)=1$, from (\ref{Q12}) it follows that the interference among the
contributions produced by the various components which take shortcuts can
explain the oscillations of the cross--section at $\theta=\pi$, which are, indeed, present in
the $\pi^+$--p elastic scattering at energy sufficiently high, as we shall see in
Subsection \ref{subse:surfacewaves}. As the energy increases the semitransparent corona around the absorbing
core gets thinner so that the effects of the shortcuts, and, accordingly, the
oscillations of the backward cross--section tend to disappear.

\section{Phenomenological analysis}
\label{se:phenomenological}

\subsection{Resonances and echoes}
\label{subse:resonances}

In the analysis of the $\pi^+$--p scattering the spin of the proton must be taken into
account. Therefore, we start with a rapid sketch of the main formulae of the scattering
amplitude in the case of spin $0$--spin $\frac{1}{2}$ collision. In particular, we have the
spin--non--flip amplitude and the spin--flip amplitude, which read, respectively:
\begin{subequations}
\label{QQ1}
\begin{eqnarray}
f(k,\theta) &=& \frac{1}{2\rmi k}\sum_{\ell=0}^\infty
\left[(\ell+1)(S^{(+)}_\ell-1)+\ell(S^{(-)}_\ell-1)\right]P_\ell(\cos\theta), \label{QQ1a} \\
g(k,\theta) &=& \frac{1}{2k}\sum_{\ell=0}^\infty
\left(S^{(+)}_\ell-S^{(-)}_\ell\right) P_\ell^{(1)}(\cos\theta), \label{QQ1b}
\end{eqnarray}
\end{subequations}
where $k$ is the c.m.s. momentum, $P_\ell^{(1)}$
is the associated Legendre function, and
\begin{subequations}
\label{QQ2}
\begin{eqnarray}
S^{(+)}_\ell &=& \exp(2\rmi\delta_{\ell,\ell+1/2}), \label{QQ2a} \\
S^{(-)}_\ell &=& \exp(2\rmi\delta_{\ell,\ell-1/2}) \label{QQ2b}
\end{eqnarray}
\end{subequations}
$\delta_{\ell,\ell\pm 1/2}$ being the phase--shift associated with the partial wave with total
angular momentum $J=\ell\pm\frac{1}{2}$, where $\frac{1}{2}$ comes from the proton spin.
As a typical example one can keep in mind the $\Delta(\frac{3}{2},\frac{3}{2})$ resonance,
where the relative angular momentum of the system is $\ell=1$ since the angular
momentum of the proton is zero, and taking into account
the spin of the proton one has the value $\frac{3}{2}$ for the total angular momentum;
analogously, summing the isotopic spin of the pion with that of the proton one obtains
for the total isotopic spin the value $\frac{3}{2}$. At higher energies one has to combine
the angular momentum of the proton (i.e., $L=2,4$) with the angular momentum
carried by the $P$--wave pion; finally, taking into account the proton spin
one gets the $J^p$ values of the first three resonances.

\begin{figure}[ht]
\begin{center}
\leavevmode
\includegraphics[width=11cm]{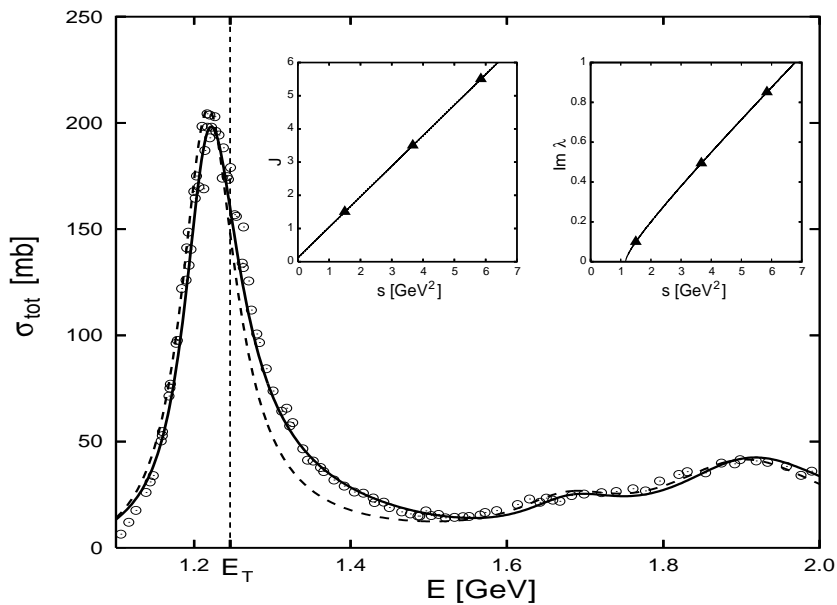}
\caption{
\label{fig:4}
Total cross--section. The experimental
data (open dots) are taken from Ref. \protect\cite{Yao}, and, for a better
visualization, only a subset of the available data have been
plotted; for the same reason no error bars are shown. The solid
line indicates the total cross--section computed by means of
Eq. (\protect\ref{QQ4}), taking into account the contributions of both the
resonance and echo poles generating $\delta_\ell^{(+)}$
(see formulae (\protect\ref{QQ5}) and (\protect\ref{QQ6})), and the pole
associated with the $\Delta(\frac{1}{2},\frac{3}{2})$ resonance.
The dashed line shows the total cross--section computed by
accounting for only the resonance poles (no echo pole).
The fitting parameters are (see Eqs. (\protect\ref{QQ9})):
$a_0=6.89 \times 10^{-1}$, $a_1= 9.2 \times 10^{-1} (\mathrm{GeV})^{-2}$,
$b_1= 9.0 \times 10^{-2} (\mathrm{GeV})^{-1}$, $b_2= 1.4 \times 10^{-1} (\mathrm{GeV})^{-2}$,
$c_0= -5.0 \times 10^{-1}$, $c_1 = 5.0 \times 10^{-1} (\mathrm{GeV})^{-2}$,
$g_0= 2.0 (\mathrm{GeV})^{-2}$, $g_1 = 3.0 (\mathrm{GeV})^{-4}$.
In the left inset $J\equiv\alpha_r+1/2$ versus $s$ is given: $J=0.92 s + 0.12$.
In the right inset $\beta_r\equiv\Imag\lambda_r$ versus $s$ is drawn. The triangles indicate the values
corresponding to the three resonances with $J^p=\frac{3}{2}^+,\frac{7}{2}^+,\frac{11}{2}^+$.
\vspace{0.5cm}
}
\end{center}
\end{figure}

The differential cross--section is given by
\beq
\label{QQ3}
\frac{d\sigma}{d\Omega} = |f|^2 + |g|^2,
\eeq
if the proton target is unpolarized, and if the Coulomb scattering is neglected.
Let us indeed note that the Sommerfeld parameter $\eta=e^2/(\hbar\nu)$ at $E=1200$ MeV
(close to the energy of the $\Delta(\frac{3}{2},\frac{3}{2})$ resonance) is of the order of $0.04$.
Next, integrating over the angles and taking into account the orthogonality
of the spherical harmonics, one obtains for the total cross--section the following expression:
\beq
\label{QQ4}
\sigma_\mathrm{tot}=\frac{2\pi}{k^2}\sum_{J,\ell}(2J+1)\sin^2\delta_{\ell,J},
\eeq
where $J=\ell\pm\frac{1}{2}$.
From formula (\ref{QQ4}) it follows that for the resonance $\Delta(\frac{3}{2},\frac{3}{2})$,
$\sigma_\mathrm{tot}=8\pi/k^2$, since $\sin^2\delta_{1,3/2}=1$ at the resonance energy.
We can now use a suitably adapted form of (\ref{Q5}) for describing
the phase--shift $\delta_{1,3/2}$, which is responsible of the $\Delta(\frac{3}{2},\frac{3}{2})$
resonance. Moreover, the same formula, at different values of $\ell$, can reproduce the
other phase--shifts $\delta_{\ell,\ell+1/2}$ ($\ell=3,5$), which generate the resonances
$\Delta(\frac{7}{2},\frac{3}{2})$ and $\Delta(\frac{11}{2},\frac{3}{2})$.
Since we interpolate the phase--shifts with odd values of $\ell$, the argument of $\sin^{-1}$
must be multiplied by the factor $\frac{1-(-1)^\ell}{2}$ which derives
from the requirement of antisymmetrization of the scattering amplitude.
Therefore, denoting for brevity $\delta_\ell^{(\pm)}\equiv\delta_{\ell,\ell\pm 1/2}$,
in the \emph{neighborhood of the resonances} we write:
\beq
\left(\delta_\ell^{(+)}\right)_r
=\sin^{-1}\left\{
\frac{1-(-1)^\ell}{2}
\frac{\beta_r(2\alpha_r+1)}
{\left\{\left[\left(\ell-\alpha_r\right)^2+\left(\beta_r\right)^2\right]
\left[\left(\ell+\alpha_r+1\right)^2+\left(\beta_r\right)^2\right]\right\}^{1/2}}\right\},
\label{QQ5}
\eeq
where the subscript '$r$' is for recalling and emphasizing that the pole located at
$\lambda_r=\alpha_r+\rmi\beta_r$ refers to resonances and lies in the first quadrant of the CAM--plane.
But formula (\ref{QQ5}) is not sufficient for representing all the features of the experimental
data. We must add also the effect of the echo indeed. To this purpose,
from formula (\ref{kk4}), and taking into account the antisymmetrization induced by the odd
values $\ell$ of the phase--shifts being considered, we have:
\beq
\left(\delta_\ell^{(+)}\right)_e
=\sin^{-1}\left\{\frac{1-(-1)^\ell}{2}
\frac{-\beta_e(2\alpha_e+1)}
{\left\{\left[\left(\ell-\alpha_e\right)^2+\left(\beta_e\right)^2\right]
\left[\left(\ell+\alpha_e+1\right)^2+\left(\beta_e\right)^2\right]\right\}^{1/2}}\right\},
\label{QQ6}
\eeq
where the subscript '$e$' is for recalling that the pole located at
$\lambda_e=\alpha_e-\rmi\beta_e$ ($\beta_e>0$) lies in the
fourth quadrant of the CAM--plane and refers to an echo.
Then, adding the two contributions (\ref{QQ5}) and (\ref{QQ6}) we obtain
an approximation for the phase--shifts, which is able to reproduce the sequence of both resonances
and echoes,
\beq
\delta_\ell^{(+)} = \left(\delta_\ell^{(+)}\right)_r + \left(\delta_\ell^{(+)}\right)_e.
\label{QQ7}
\eeq
In order to fit the total cross--section over a large interval of energy (up to $2\,\mathrm{GeV}$; see Fig. \ref{fig:4}),
we must consider also the contribution coming from the resonance
$\Delta(\frac{1}{2},\frac{3}{2})$ with $J^P=\frac{1}{2}^-$ and $E\simeq 1620\,\mathrm{MeV}$.
In view of its negative parity this resonance does not belong to the family of resonances being considered,
whose parity is even, and, consequently, we may treat it separately.
To this end we can add in formula (\ref{QQ4}) an expression of the phase--shift associated to the
$\Delta(\frac{1}{2},\frac{3}{2})$ strictly analogous
to formula (\ref{QQ5}); moreover, the corresponding echo term  can be neglected in view of its small effect.
Now, returning to formulae (\ref{QQ5}) and (\ref{QQ6}),
the functions $\alpha_r$, $\beta_r$, $\alpha_e$, $\beta_e$ can be parameterized as follows (see also Refs. \cite{DeMicheli1,DeMicheli3}):
\begin{subequations}
\label{QQ9}
\begin{eqnarray}
\alpha_r &=& a_0 + a_1 (E^2-E_0^2), \label{QQ9a} \\
\beta_r  &=& b_1 (E^2-E_0^2)^{1/2} + b_2 (E^2-E_0^2), \label{QQ9b} \\
\alpha_e &=& c_0 + c_1 (E^2-E_0^2), \label{QQ9c} \\
\beta_e  &=& g_0 (E^2-E_0^2) + g_1 (E^2-E_0^2). \label{QQ9d}
\end{eqnarray}
\end{subequations}
where $E$ is the energy in the c.m.s., and $E_0$ is the rest mass
of the $\pi^+$--p system. Finally, substituting in formula
(\ref{QQ4}) the values of $\delta_{\ell,J}$ obtained by formulae
(\ref{QQ5}), (\ref{QQ6}), (\ref{QQ7}) and (\ref{QQ9}), the
experimental total cross--section can be fitted.
The result is the solid line shown in Fig. \ref{fig:4}.
From this fit we observe clearly two resonances
of even parity, whose $J^p$ values are $J^p=\frac{3}{2}^+$, and $\frac{7}{2}^+$.
A third resonance of even parity, with $J^p=\frac{11}{2}^+$, is not visible
but can be extrapolated by computing $(\delta^{(+)}_{(\ell=5)})_{r,e}$ with the parameters
obtained from the analysis of the $\Delta(\frac{3}{2},\frac{3}{2})$ and
$\Delta(\frac{7}{2},\frac{3}{2})$ resonances. The numerical values obtained in this way are
summarized in Table \ref{tab:1}. Plotting $J=\alpha_r+1/2$ ($\alpha_r\equiv\Real\lambda_r$)
versus $s$, we obtain the straight line displayed in the first inset of Fig. \ref{fig:4}. Particularly relevant is the
value obtained for $a_1$, i.e., $a_1 \simeq 1/\mathrm{(GeV)}^2$
(see formula (\ref{QQ9a})), which gives the slope of the linear trajectory.
We thus obtain a \emph{phenomenological evidence of a vibrational--like spectrum generated by the
$\fsp(3,\R)$ algebra, as it has been derived, at non--relativistic level, in Subsection \ref{subse:vibrational}}.
The second inset of Fig. \ref{fig:4} shows the plot of $\beta_r\equiv\Imag\lambda$ against $s$ along with the values
attained at the three resonances;
this behavior correctly describes the phenomenological observation that the width $\Gamma$ of the resonances
increases with the energy (see Table \ref{tab:1} for the numerical values).
The dashed line in Fig. \ref{fig:4} displays the fit of the total cross--section computed without the
echo terms (see Eqs. (\ref{QQ9}c and d)), and shows with no ambiguity the necessity of introducing
a pole singularity in the fourth quadrant of the CAM--plane, as it has been explained
in Subsection \ref{subse:surface}. In the case of the $\Delta(\frac{3}{2},\frac{3}{2})$ resonance, where
the effect is particularly evident, a naive semiclassical argument can support the interpretation of the distortion
of the symmetric bell--shaped peak,
which is clearly exhibited by the experimental data, in terms of the composite nature of the interacting particles.
If we denote by $R_{\pi\mathrm{p}}$ the distance between the pion and the proton, supposed at rest, then the
impulse of the incoming pion in the lab frame is $p_\mathrm{lab}^\pi\sim\sqrt{2}\hbar/R_{\pi\mathrm{p}}$, since $\ell=1$.
Now, if we set $R_{\pi\mathrm{p}}$ as the distance at which the two particles \emph{``get in contact"}, which can be thought
of the order of the proton radius, then the corresponding $p_\mathrm{lab}^\pi$ yields an estimate
of the least pion impulse at which the fermionic nature of the constituents of the colliding particles enter the picture
in the collision process. Then, using $R_{\pi\mathrm{p}}=R_\mathrm{proton} \simeq 0.87 \,\mathrm{fm}$ we have
$p_\mathrm{lab}^\pi\sim 320\,\mathrm{GeV/c}$, which corresponds to the center of mass energy $E_T\sim 1246\,\mathrm{MeV}$,
which is indicated in Fig. \ref{fig:4}.

\begin{table}[tb]
\caption{
\label{tab:1}
$\pi^+$--p elastic scattering: Analysis of the resonances.
The \emph{purely resonant mass} in the fourth column gives the mass of the resonance computed by
means of Eq. (\protect\ref{QQ9a}) without the contribution coming from the echo pole (see Eq. (\protect\ref{QQ9c})).
The \emph{purely resonant width} $\Gamma_R$ indicates the width of the resonance peak
computed without the echo contribution, while the \emph{total} width $\Gamma$ stands for
the width of the resonance peak accounting also for the echo term.
}
{\scriptsize
\begin{center}
\leavevmode
\begin{tabular}{ccccccc}
\\ \hline \\
$J^P$ & Mass [MeV] & Mass [MeV] & Mass [MeV] & $\Gamma_R$ [MeV] & $\Gamma$ [MeV] & $\Gamma$ [MeV] \\[+3pt]
 & (present work) & (Ref. \protect\cite{Yao}) & Resonant & Resonant & Total & (Ref. \protect\cite{Yao}) \\[+5pt] \hline \\
$\frac{3}{2}^+$&1232.8&$1230 - 1233$ & $1224$ & $93$ & 115 & $116 - 120$ \\[+5pt]
$\frac{7}{2}^+$&1951&$1915 - 1950$ & $1916$ & $293$ & 308 & $235 - 335$ \\[+5pt]
$\frac{11}{2}^+$&2463&$2300 - 2500$ & $2418$ & $397$ & 410 & $300 - 500$
\\[+5pt] \hline
\end{tabular}
\end{center}
}
\vspace{0.5cm}
\end{table}

\begin{remark}
\label{rem:3}
\rm
We have obtained the fits shown in Fig. \ref{fig:4} by using formulae (\ref{QQ5}) and (\ref{QQ6})
for the phase--shifts. In these formulae the expression $[\pm\beta_{r,e}(2\alpha_{r,e}+1)]$
has been obtained by making use of the unitarity condition (see (\ref{kk3})).
This condition holds true only up to a certain value of energy. It is certainly admissible
in the energy range including the $\Delta(\frac{3}{2},\frac{3}{2})$ resonance, say, up to
$E \sim 1.5 \,\mathrm{GeV}$. Therefore, we can be sure for what concerns the
effect of the echo and its explanation by the introduction of a pole in the
fourth quadrant of the CAM--plane, as illustrated in Subsection \ref{subse:surface} and depicted in
Fig. \ref{fig:4}. But, returning to Fig. \ref{fig:4}, we have obtained
a good agreement of our fitting formulae with the experimental data up to
$E \sim 2.0 \,\mathrm{GeV}$, in spite of the fact that the unitarity condition
is certainly violated for $1.5 \,\mathrm{GeV}\lesssim E \lesssim 2.0 \,\mathrm{GeV}$.
For higher values of energy our fits break down as expected. This fact can be tentatively
explained conjecturing that our formulae still represent an admissible approximation
also beyond the region where the unitarity condition is strictly valid.
\end{remark}

\subsection{Surface waves and Sommerfeld poles}
\label{subse:surfacewaves}

Increasing the energy, at a first cursory examination of the data,
a general picture seems to emerge: ``\emph{The amplitudes are dominated by
diffractive and peripheral contributions}'' \cite{Hendry1,Hendry2}.
For what concerns the resonances H\"ohler writes \cite[pag. S206]{Hohler}: ``\emph{If the resonances are ordered
according to the shapes of their Argand plots, one finds a continuous transition
from textbook--type resonances to tiny wiggles superimposed on a large background.}''
Furthermore, Hendry \cite{Hendry1} remarks about resonances that: ``\emph{The background
for a dominant loop tends to be small for energies below the resonance, but grows strongly
above the resonance, and quickly swamps it. In terms of diffractive and peripheral
picture what is happening is that, as the energy is increased through the resonant
region for a particular partial wave, the resonating piece of the partial wave
gets buried in a growing diffractive peak.}'' Last but not least,
at $p_\mathrm{cm} \sim 10$ GeV/c a very large number of partial waves (at least $25$) need to be
included and this makes the search problem rather cumbersome. In spite of these
difficulties and ambiguities, Hendry \cite{Hendry1} concludes that
there is a good evidence for resonances up to masses of about $4$ GeV and spins
$\frac{21}{2}$. We prefer to follow another approach and investigate up to what extent
the diffractive effects can be explained and described by Sommerfeld's poles.

\begin{figure}
\begin{center}
\leavevmode
\includegraphics[width=12cm]{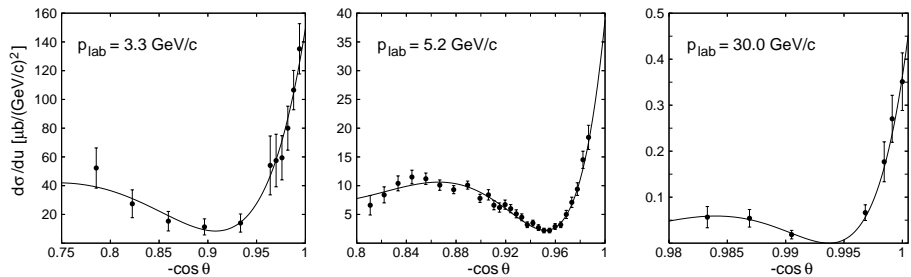}
\caption{
\label{fig:5}
Differential cross--section vs. $(-\cos\theta)$ at $p_\mathrm{lab}=3.30, 5.20, 30.00 \,\mathrm{GeV/c}$.
The solid lines show the differential cross--sections obtained by fitting the experimental data with
the function $(\rmd\sigma/\rmd u) = B_0 |P_\lambda(-\cos\theta)|^2$,
$B_0$ and $\lambda$ being the fitting parameters.
See Table {\protect\ref{tab:2}} for the summary of the numerical results obtained
for momenta ranging in $2.85 \,\mathrm{GeV/c} \leqslant p_\mathrm{lab} \leqslant 70 \,\mathrm{GeV/c}$.
\vspace{0.5cm}
}
\end{center}
\end{figure}

We may start from an appropriate extension of formula (\ref{KK})
to account for both the spin--non--flip and the spin--flip amplitudes:
\beq
\label{SS1}
\frac{\rmd\sigma}{\rmd u} \simeq
B_0 |P_\lambda(-\cos\theta)|^2 + B_1 |P_\lambda^{(1)}(-\cos\theta)|^2,
\eeq
$u$ denoting the appropriate Mandelstam variable. The term $B_1 |P_\lambda^{(1)}(-\cos\theta)|^2$
in formula (\ref{SS1}) gives the spin--flip contribution to the differential cross--section.
Let us recall, first of all, that formula (\ref{SS1}) (as formula (\ref{KK}))
describes the differential cross--section only in the backward direction. Then one
notes that by using (\ref{SS1}) two types of fits can be performed: (i) at fixed energy;
(ii) at fixed angle, i.e., $\theta=\pi$.
In Fig. \ref{fig:5} we present fits of the differential cross--section in the backward angular region at various fixed energies.
In addition, Table \ref{tab:2} summarizes the
numerical results obtained from the analysis of the backward differential
cross--section for pion laboratory momenta ranging in the interval
$2.85 \,\mathrm{GeV/c} \leqslant p_\mathrm{lab} \leqslant 70 \,\mathrm{GeV/c}$.
\begin{table}[tb]
\caption{\label{tab:2} $\pi^+$--p elastic scattering: Analysis of the surface waves.
Summary of the data resulting from the fits of the differential cross--section
at backward angles by means of formula (\protect\ref{SS1}) with $B_1=0$.
The angular data, taken from the reference given in the sixth column, have been fitted in the angular
range given in the fifth column (see text).
}
{\scriptsize
\begin{center}
\leavevmode
\begin{tabular}{cccccc}
\\ \hline \\[+3pt]
$p_\mathrm{lab}$ & $B_0$                    & $\Real\lambda$ & $\Imag\lambda$ & $(-\cos\theta)$ range & Reference                     \\[+3pt]
$\mathrm{GeV/c}$ & $\mu\mathrm{b}/(\mathrm{GeV/c})^2$ & ---         & ---         & ---                  & ---                        \\[+3pt]
\hline\\
2.85           & $495.30 \pm 23.01$         & $4.01 \pm 0.10$& $0.72 \pm 0.08$& [0.75:1]              & \protect\cite{Banaigs,Baker1} \\[+3pt]
3.30           & $149.06 \pm 8.18$          & $4.93 \pm 0.16$& $1.00 \pm 0.11$& [0.75:1]              & \protect\cite{Banaigs,Baker1} \\[+3pt]
3.55           & $131.91 \pm 7.02$          & $4.81 \pm 0.16$& $0.91 \pm 0.12$& [0.75:1]              & \protect\cite{Banaigs,Baker1} \\[+3pt]
5.20           & $38.82 \pm 1.88$           & $7.07 \pm 0.06$& $1.41 \pm 0.06$& [0.80:1]              & \protect\cite{Baker2}         \\[+3pt]
5.91           & $36.88 \pm 1.79$           & $7.63 \pm 0.09$& $0.53 \pm 0.18$& [0.80:1]              & \protect\cite{Owen}           \\[+3pt]
7.00           & $15.93 \pm 1.86$           & $8.79 \pm 0.15$& $1.52 \pm 0.19$& [0.85:1]              & \protect\cite{Baker2}         \\[+3pt]
9.85           & $8.15 \pm 0.79$            & $10.40\pm 0.26$& $1.08 \pm 0.23$& [0.95:1]              & \protect\cite{Owen}           \\[+3pt]
13.73          & $3.65 \pm 1.07$            & $12.26\pm 1.46$& $\sim 0.0$     & [0.97:1]              & \protect\cite{Owen}           \\[+3pt]
30.00          & $(3.62 \pm 0.40)\times 10^{-1}$ & $20.91 \pm 0.68$& $1.37 \pm 1.71$ & [0.98:1]       & \protect\cite{Baker3}         \\[+3pt]
50.00          & $(9.48 \pm 1.56)\times 10^{-2}$ & $28.20 \pm 1.65$& $3.31 \pm 1.97$& [0.99:1]        & \protect\cite{Baker3}         \\[+3pt]
70.00          & $(3.41 \pm 0.33)\times 10^{-2}$ & $32.82 \pm 1.86$& $4.98 \pm 1.91$& [0.99:1]        & \protect\cite{Baker3}         \\[+3pt]
\hline
\end{tabular}
\end{center}
}
\vspace{0.5cm}
\end{table}
The first observation emerging from these fits is that $B_1$ (see formula (\ref{SS1}))
is negligible compared to $B_0$. Then we take $\Real\lambda\equiv\alpha$,
$\Imag\lambda\equiv\beta$, and $B_0$ as fitting parameters. We are thus able
to relate the values of $\lambda=\alpha+\rmi\beta$ to the location of Sommerfeld's poles:
more precisely, to the root of $H_\nu^{(1)}(kR)=0$ closest to the real axis (see (\ref{Q11})).
Next, we observe that these angular fits are successful if we limit
ourselves to consider an extremely backward angular region, whose range has a decreasing spread
at increasing energy. A possible explanation of this phenomenology is that,
as the energy increases, the distance between two consecutive zeros of $H_\nu^{(1)}(kR)$ tends to zero,
and that the arguments of the zeros tend to $\pi/2$, although their real part tends to infinity \cite{Keller2,Magnus}.
Therefore it appears quite reasonable to conjecture that, at increasing energy,
the zero of $H_\nu^{(1)}(kR)$ closest to the real axis is not sufficient anymore to fit the data,
but other zeros of $H_\nu^{(1)}(kR)$ come into play.
\begin{figure}
\begin{center}
\leavevmode
\includegraphics[width=11cm]{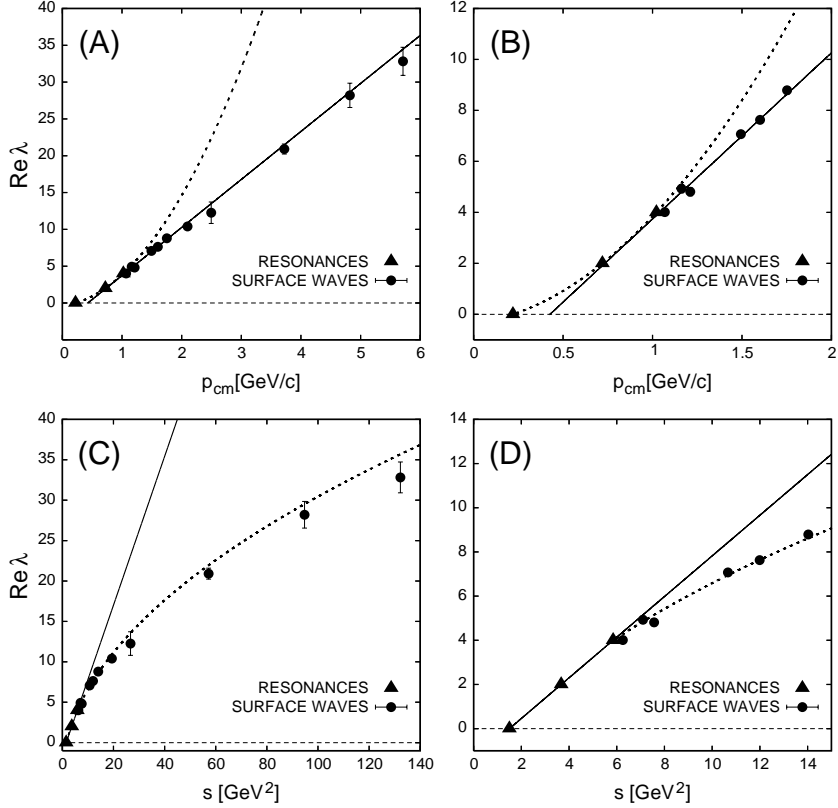}
\caption{
\label{fig:6}
$\Real\lambda$ obtained from the analysis of resonances and surface waves.
(A) $\Real\lambda$ vs. $p_\mathrm{cm}$. The dots indicate the values of $\Real\lambda$ obtained
from the analysis of the differential cross--section (see text, Table \protect\ref{tab:2}, and the
legend of Fig. \protect\ref{fig:5}). The triangles indicate the values of $\Real\lambda$ obtained
from the analysis of the resonances (see text, Table \protect\ref{tab:1}, and the legend of Fig. \protect\ref{fig:4}). These
values have been lowered by 1 in order to account for the unity pion angular momentum (see Remark \protect\ref{rem:4}).
The values of $\Real\lambda$ obtained from the analysis of the differential cross--section
have been fitted with a straight line $\Real\lambda(p_\mathrm{cm}) = r_0 \,(p_\mathrm{cm}-p_\mathrm{cm}^*)$
(solid line). The fitting parameters are:
$r_0 = 6.52 \pm 0.19 \, (\mathrm{GeV/c})^{-1}$; $p_\mathrm{cm}^* = (4.3 \pm 0.6)\times 10^{-1}\,\mathrm{GeV/c}$.
From the value of the parameter $r_0$ it results a pion-proton effective interaction radius $R \sim 1.28 \,\mathrm{fm}$
(see Eq. (\protect\ref{Q11})). The dashed line indicates the curve on which the real part of the
angular momentum associated with the resonances lie (see Eq. (\protect\ref{QQ9a}) and the legend
of Fig. \protect\ref{fig:4}).
(B) Zoom of the panel (A) in the momentum range $0<p_\mathrm{cm}<2 \, \mathrm{GeV/c}$ for better visualization of the
transition region.
(C) $\Real\lambda$ vs. $s$. As in the previous panels, the triangles denote the values of $\Real\lambda$
(lowered by 1) associated with the resonances with $J^p = \frac{3}{2}^+,\frac{7}{2}^+,\frac{11}{2}^+$. The straight
line shows the related Regge trajectory
(see Eq. (\protect\ref{Q11}) and the legend of Fig. \protect\ref{fig:4}).
The dashed line shows the fit of the data coming from the analysis of the surface waves
(filled dots) with the function $(d_0 + d_1 \sqrt{s})$. The fitting parameters are:
$d_0 = -4.45 \pm 0.41$; $d_1 = (3.49 \pm 0.13) \, (\mathrm{GeV})^{-1}$.
(D) Zoom of the panel (C) in the $s$--range $0<s<15 \,(\mathrm{GeV})^2$. The dashed line shows
the deviation of the real part of the angular momentum associated with the surface
waves from the linear Regge trajectory.
}
\end{center}
\end{figure}
The values of $\Real\lambda$ coming from the analysis of the differential cross--section in the backward angular region
are displayed in Figs. \ref{fig:6}A and B (filled dots) as function of $p_\mathrm{cm}$ ($p_\mathrm{cm}=\hbar k$).
They clearly appear to lie on a straight line in very good agreement
with Sommerfeld's formula (\ref{Q11}). This linear behavior is very neat, and is primarily
related to the location of the dip in the backward differential cross--section. In Figs. \ref{fig:6}C and D
the same data are plotted against $s$ (instead of $p_\mathrm{cm}$), and we obtain a square--root dependence,
i.e., $\Real\lambda\sim\sqrt{s}$. Moreover, in the same figure, the values of $\Real\lambda$ associated
with the sequence of resonances are also given (filled triangles).
From Fig. \ref{fig:6} it emerges clearly that \emph{it is impossible,
plotting $\Real\lambda$ against $p_\mathrm{cm}$ (or, equivalently, against $s$) to locate both the resonances
and the surface waves on the same trajectory, i.e., on a single straight line}.
In particular, in disagreement with Hendry's analysis, we see that it is impossible to locate
the resonances (in particular the $\Delta(\frac{3}{2},\frac{3}{2})$ resonance)
on a straight line plotting $\Real\lambda$ versus $p_\mathrm{cm}$ (see Figs. \ref{fig:6}A and B). It follows
that \emph{resonances and surface waves correspond to two different classes of poles indeed}:
the resonances can be described by poles nearly parallel to the real axis of the CAM--plane;
the surface waves by poles nearly parallel to the imaginary axis of the CAM--plane.
It appears, however, that in the neighborhood of $p_\mathrm{cm}\sim 1.2\,\mathrm{GeV/c}$
a transition occurs between these two classes of poles.

\begin{remark}
\label{rem:4}
\rm
In Fig. \ref{fig:6} the values of $\Real\lambda$ corresponding to the resonances have been
lowered by 1, which is precisely the value, in $\hbar$ units, of the angular momentum
of the incoming pion. This subtraction can be justified by observing that in the transition
from resonances to surface waves, i.e., from quantum to semiclassical dynamics, the unity angular
momentum of the incoming pion smears out in a continuous set of values.
Therefore by this artful lowering of 1 ($\hbar$) the connection between resonances and surface
waves merges with clear evidence.
\end{remark}

\begin{figure}
\begin{center}
\leavevmode
\includegraphics[width=7cm]{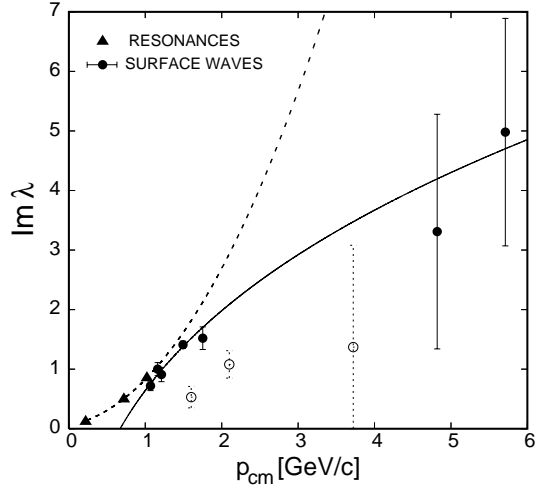}
\caption{
\label{fig:7}
$\Imag\lambda$ obtained from the analysis of resonances and surface waves.
The dots indicate the values of $\Imag\lambda$ obtained from the analysis of the differential
cross--section (see text, Table \protect\ref{tab:2}, and the legend of Fig. \protect\ref{fig:5}).
The triangles indicate the values of $\Imag\lambda$ associated with
the resonances (see Eq. (\protect\ref{QQ9b}) and the legend of Fig. \protect\ref{fig:4}).
The dashed line displays the fit of the data coming from the analysis of the
surface waves with the function $[h_1 (p_\mathrm{cm})^{1/3} + h_0]$
(see Eq. (\protect\ref{Q11})). The fitting parameters are: $h_0 = -4.51 \pm 0.54$;
$h_1 = (5.15 \pm 0.49) \, (\mathrm{GeV/c})^{-1/3}$. From the value of $h_1$ it follows
from Eq. (\protect\ref{Q11}) the interaction radius $R \sim 6 \, \mathrm{fm}$. The values of $\Imag\lambda$
resulting from the analysis of the experimental data at $p_\mathrm{lab} = 5.91, 9.85, 13.73 \, \mathrm{GeV/c}$,
taken from Ref. \protect\cite{Owen}, and at $p_\mathrm{lab} = 30 \,\mathrm{GeV/c}$ from Ref. \protect\cite{Baker3},
are shown in ghost form, and have not been included in the fitting procedure.
\vspace{0.5cm}
}
\end{center}
\end{figure}

In Fig. \ref{fig:7} the plot of $\Imag\lambda\equiv\beta$ versus $p_\mathrm{cm}$ ($p_\mathrm{cm}\equiv \hbar k$) is given;
Although these data must be taken with care in view of the large errors and of the presence of a few outliers,
this figure shows that the values of $\Imag\lambda$ coming from the analysis of the
surface waves may belong to a curve whose $k$--dependence is of the form
$\beta\sim (kR)^{1/3}$, in accord with Sommerfeld's pole formula (\ref{Q11}).
But we must pay a price for this since the value obtained for $R$ is of the order
of $6\,\mathrm{fm}$. Interestingly, as we have seen in Fig. \ref{fig:6}, also the
behavior of $\Imag\lambda$ indicates a transition between Regge and Sommerfeld poles in a
neighborhood of $p_\mathrm{cm}\sim 1.2\,\mathrm{GeV/c}$ (see Fig. \ref{fig:6}B).

\begin{figure}
\begin{center}
\leavevmode
\includegraphics[width=11cm]{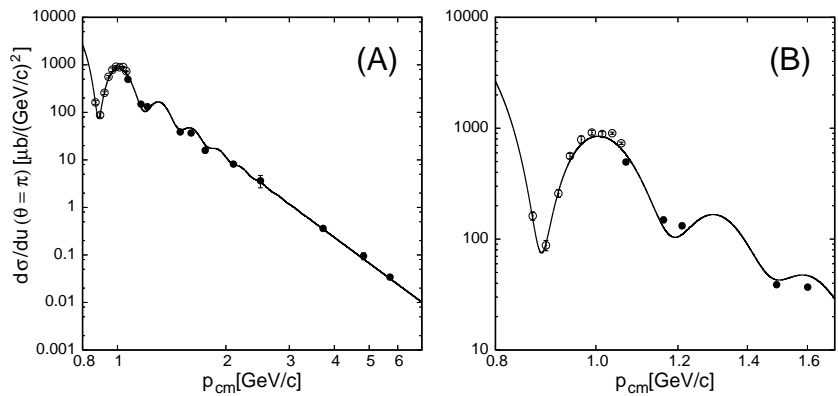}
\caption{
\label{fig:8}
Differential cross--section at $\theta = \pi$ vs. center of mass momentum.
(A) The filled dots represent the values of the parameter $B_0$,
i.e., the differential cross--section extrapolated at $\theta=\pi$, obtained from the analysis of the angular
distribution (see Fig. \protect\ref{fig:5}); the numerical values are summarized in Table \protect\ref{tab:2}.
The open dots denote the differential cross--section extrapolated at $\theta=\pi$, obtained from the
analysis of the angular distributions given in Ref. \protect\cite{Lennox}. For the sake of consistency with the analysis
performed at higher momenta, the experimental angular data given in
Ref. \protect\cite{Lennox} have been fitted with the function $(\rmd\sigma/\rmd u) = B_0 |P_\lambda(-\cos\theta)|^2$
(as in the analysis shown in Fig. \protect\ref{fig:5}). However in this case, in view of the limited number
of angular data available at each momentum, we used only $B_0$ as a fitting parameter.
At each momentum, we have set $\Real\lambda(p_\mathrm{cm})$ at the value extrapolated from
the analysis at higher momenta (see the straight line in Fig. \protect\ref{fig:6}A), and $\Imag\lambda=0$.
These values of differential cross--section at $\theta=\pi$ differ by a few percent from those
given in Ref. \protect\cite{Lennox}, which have been extrapolated by using different methods.
The solid line represents the fit of the data with the function given in (\protect\ref{ampli});
the fitting parameters are:
$p_0= (3.09\pm 0.03) \,\mathrm{GeV/c}$;
$c_1 = -5.61 \pm 0.07$;
$c_2 = (-5.55 \pm 1.64) \,\mu\mathrm{b}/(\mathrm{GeV/c})^2$;
$c_3 = (2.01 \pm 0.33) \,(\mathrm{GeV/c})^{-1}$;
$\omega = (21.18 \pm 0.04) \,(\mathrm{GeV/c})^{-1}$.
(B) Zoom of panel (A) in the range $0.8\,\mathrm{GeV/c} < p_\mathrm{cm} < 1.7\,\mathrm{GeV/c}$ for better visualization of the
oscillatory behavior at low momenta.
\vspace{0.5cm}
}
\end{center}
\end{figure}

In Fig. \ref{fig:8} the data of the differential cross--section at $\theta=\pi$ (see the term $B_0$
in Eq. (\ref{SS1}), and Table \ref{tab:2}) are given as function of $p_\mathrm{cm}$.
From a first cursory examination of the data we observe at low
$p_\mathrm{cm}$ an anomalous large peak which resembles that encountered in nuclear
physics (e.g., in $\alpha$--$^{40}\mathrm{Ca}$ elastic scattering) giving rise to the
phenomenon called ``ALAS'' (anomalous large angle scattering). Moreover, an oscillating
pattern seems to be superimposed over an inverse power trend (notice the bilogarithmic scale).
As the momentum increases the amplitude
of the oscillations decreases and tends to zero at large momentum. Also the value of the backward
peak is rapidly decreasing for higher values of the center of mass momentum. A first
qualitative explanation of these phenomena can be given by following the model presented in
Subsection \ref{subse:surface} (see formula (\ref{Q12})). In the transition region
from resonances to surface waves the target appears as an opaque ball at the center surrounded
by a semitransparent corona. The hitting rays, which are not absorbed and pass through
the corona, focus at backward angles, and produce the anomalous peak. When the
momentum increases, the radius of the opaque core increases too, and the backward peak, after some oscillations
of decreasing amplitude, diminishes as the semitransparent corona becomes thinner and the
blackbody limit is reached. A phenomenological quantitative fit of the data can be performed
by means of the following formula:
\beq
\label{ampli}
\left(\frac{\rmd\sigma}{\rmd u}\right)_{\theta=\pi}
= \left(\frac{p_\mathrm{cm}}{p_0}\right)^{c_1}\left[1+c_2 e^{-c_3\, p_\mathrm{cm}}\cos(\omega p_\mathrm{cm})\right],
\eeq
where $p_0$, $c_i$ ($i=1,2,3$), and $\omega$ are determined through the fit of the
experimental data, and their values are given in the legend of Fig. \ref{fig:8}. Formula
(\ref{ampli}) describes the oscillating decrease of the backward differential cross--section
as an inverse power of $p_\mathrm{cm}$ ($c_1 < 0$) with a superimposed oscillating pattern whose amplitude
decreases exponentially with the momentum. Let us note that a backward peak in the differential
cross--section can also be described by the \emph{baryon exchange mechanism} in the sense of the
conventional Regge--pole exchange theory. It is remarkable to note that by the use of this theory
one obtains a differential cross--section at $\theta=\pi$ decreasing as the inverse power of $p$
with an exponent which is very similar to that obtained by fitting the experimental data with
formula (\ref{ampli}) \cite{Lyubimov}.

The other relevant feature of the differential cross--section is the forward diffraction peak.
But, at $\theta=0$, the function $P_\lambda(-\cos\theta)$ presents a logarithmic singularity
of the following type \cite{Sommerfeld}:
\beq
\label{SS4}
P_\lambda(-\cos\theta) \rightarrow \frac{\sin\pi\lambda}{\pi}\log\theta^2,
\eeq
and, consequently, the model elaborated in Subsection \ref{subse:surface} cannot be used.
In the forward scattering the whole sequence of Sommerfeld's poles enter the game since the surface
waves describe a small arc of meridian circumference. Furthermore the contribution of the
background integral cannot be neglected. One is then led to the ambiguous compensation between two divergent
terms: the infinite due to the surface waves at $\theta=0$ and the background integral. These
two terms should compensate in order to obtain a finite and regular amplitude at $\theta=0$.
The resulting amplitude is the forward diffractive peak.
However it is out of our purposes to analyze here the scattering in the forward region, which has been
studied, as well--known, by using several different methods (for a very recent review on this topic see Ref. \cite{Block}).
Among the others, one can refer once again to the method based on the
\emph{Regge pole exchange}.
In view also of these considerations, let us finally remark that our analysis leading
to two different classes of poles, Regge's and Sommerfeld's poles in the direct channel,
are very well consistent with the standard method of Regge pole exchange.

\section{Conclusions}
\label{se:conclusions}

(1) In agreement with Hendry's analysis we show that in the elastic $\pi^+$--p scattering
only the first few resonances, precisely the first three, whose $J^p$ values are
$J^p=\frac{3}{2}^+,\frac{7}{2}^+,\frac{11}{2}^+$, lie on a straight line trajectory of a pole
in the CAM--plane having the form: $J^p=\alpha_0+\alpha' m^2$, $\alpha'\simeq 1/\mathrm{(GeV)}^2$,
which is the standard expression of Regge pole trajectory. The conventional picture of linear rising
Regge trajectories with universal slope does not hold. \\
(2) In disagreement with Hendry's analysis we prove that, concerning these first three resonances,
it is not possible to obtain a straight line behavior if we plot $J$ as a function of the center of mass
momentum $k$. \\
(3) Beside resonances one must take into account also echoes in order to fit
the total cross--section. The advantage of using pole singularities lying in the CAM--plane
consists in the possibility of describing the echoes by poles in the fourth quadrant
of the CAM--plane, instead of introducing the scattering by an impenetrable sphere. \\
(4) In agreement with our theoretical analysis, we show that the first three resonances can be associated,
in a non--relativistic model, to a vibrational--like spectrum generated by the $\fsp(3,\R)$ algebra. \\
(5) At higher energies the amplitudes are dominated by diffractive effects, whose most peculiar
features are the creeping waves. \\
(6) Resonances and creeping waves cannot be described by the same class of poles. The
resonances can be described by a class of poles close to the real positive semi--axis of the
CAM--plane: Regge poles. The surface waves can be described by a class of poles lying in the
first quadrant of the CAM--plane, which are nearly parallel to the imaginary axis:
Sommerfeld's poles. \\
(7) At high energy, where the surface waves are dominant, we may fit the differential
cross--section at backward angles by the formula $(\rmd\sigma/\rmd u)\simeq B_0 |P_{\alpha+\rmi\beta}(-\cos\theta)|^2$,
$\alpha$ and $\beta$ giving the location of the Sommerfeld pole which is closest to the
real axis of the CAM--plane. After the well pronounced peak at $\theta=\pi$, there is a dip, whose
location is related to $\alpha\equiv\Real\lambda$. Plotting $\alpha$ versus $k$ we obtain a straight
line in good agreement with Sommerfeld's formula which gives $\Real\lambda +\frac{1}{2} \simeq kR$.

\end{document}